\documentclass[10pt,twocolumn,english,prd,superscriptaddress,nofootinbib,preprintnumbers,showpacs,floatfix]{revtex4-1}
\usepackage[utf8]{inputenc}
\usepackage{amsmath}
\usepackage{amsfonts}
\usepackage{amssymb}
\usepackage{graphicx}
\graphicspath{ {figures/} }
\usepackage[usenames,dvipsnames]{xcolor}
\usepackage{hyperref}   
\hypersetup{colorlinks=true, 
citecolor=MidnightBlue, linkcolor=MidnightBlue, urlcolor=Cyan}
\usepackage{tikz}
\usepackage{verbatim} % comment
\definecolor{purple}{rgb}{1,0,1}
\definecolor{lime}{HTML}{A6CE39} % needs xcolor

\newcommand{\orcidicon}{%
	\begin{tikzpicture}
	\draw[lime, fill=lime] (0,0) 
		circle [radius=0.16] 
		node[white] {{\fontfamily{qag}\selectfont \tiny ID}};
	\draw[white, fill=white] (-0.0625,0.095) 
		circle [radius=0.007];
	\end{tikzpicture}	\hspace{-2mm}
}
\newcommand\orcidFrancisco{{\href{https://orcid.org/0000-0002-9388-8373}{\orcidicon}}}

\newcommand\orcidManuel{{\href{https://orcid.org/0000-0001-8586-0285}{\orcidicon}}}

\newcommand\orcidTarciso{{\href{https://orcid.org/0009-0007-0450-2672}{\orcidicon}}}
\begin{document}
%========================================================
%\title{Conformal Killing gravity: black holes, regular black holes, and black bounces}
\title{Black bounces in conformal Killing gravity}

%========================================================
	\author{Jos\'{e} Tarciso S. S. Junior\orcidTarciso\!\!}
 \email{tarcisojunior17@gmail.com}
\affiliation{Faculdade de F\'{\i}sica, Programa de P\'{o}s-Gradua\c{c}\~{a}o em 
F\'isica, Universidade Federal do 
 Par\'{a},  66075-110, Bel\'{e}m, Par\'{a}, Brazil}

%=================================================================
	\author{Francisco S. N. Lobo\orcidFrancisco\!\!} \email{fslobo@ciencias.ulisboa.pt}
\affiliation{Instituto de Astrof\'{i}sica e Ci\^{e}ncias do Espa\c{c}o, Faculdade de Ci\^{e}ncias da Universidade de Lisboa, Edifício C8, Campo Grande, P-1749-016 Lisbon, Portugal}
\affiliation{Departamento de F\'{i}sica, Faculdade de Ci\^{e}ncias da Universidade de Lisboa, Edif\'{i}cio C8, Campo Grande, P-1749-016 Lisbon, Portugal}

%=================================================================
	\author{Manuel E. Rodrigues\orcidManuel\!\!}
	\email{esialg@gmail.com}
	\affiliation{Faculdade de F\'{\i}sica, Programa de P\'{o}s-Gradua\c{c}\~{a}o em 
F\'isica, Universidade Federal do 
 Par\'{a},  66075-110, Bel\'{e}m, Par\'{a}, Brazil}
\affiliation{Faculdade de Ci\^{e}ncias Exatas e Tecnologia, 
Universidade Federal do Par\'{a}\\
Campus Universit\'{a}rio de Abaetetuba, 68440-000, Abaetetuba, Par\'{a}, 
Brazil}

%-----------------------------------------------------------------
\date{\LaTeX-ed \today}
%%%%%%%%%%%%%%
%========================================================
\begin{abstract}
%========================================================
In this work, we analyse black bounce solutions in the recently proposed ``Conformal Killing gravity'' (CKG), by coupling the theory to nonlinear electrodynamics (NLED) and scalar fields. 
The original motivation of the theory was essentially to fulfil specific criteria that are absent in existing gravitational theories, namely, to obtain the cosmological constant as an integration constant, derive the energy-momentum conservation law as a consequence of the gravitational field equations, rather than assuming it, and not necessarily considering conformally flat metrics as vacuum solutions. In this work, we extend the static and spherically symmetric solutions obtained in the literature, and explore the possibility of black bounces in CKG, coupled to NLED and scalar fields. We find novel NLED Lagrangian densities and scalar potentials, and extend the class of black bounce solutions found in the literature. Furthermore, within black bounce geometries, we find generalizations of the Bardeen-type and Simpson-Visser geometries and explore the regularity conditions of the solutions. 
\end{abstract}
%========================================================
%=================================================================
\maketitle
%=================================================================
\def\HMS{{\scriptscriptstyle{\rm HMS}}}
%========================================================
%\bigskip
%\hrule
%\tableofcontents
%\bigskip
%\hrule
%========================================================
%\parindent0pt
%\parskip7pt
%========================================================

%%%%%%%%%%%%%%%%%%%%%%%

%%%%%%%%%%%%%%%%%%%%%%%%%%%%%%%%%%%%%%%%%%%%%%%%%
\section{Introduction}\label{sec1}
%%%%%%%%%%%%%%%%%%%%%%%%%%%%%%%%%%%%%%%%%%%%%%%%%

Recently, Harada proposed a novel modified gravitational theory \cite{Harada}, denoted as Conformal Killing Gravity (CKG) \cite{Mantica:2023stl}, that satisfies several theoretical criteria for gravitational theories beyond General Relativity (GR), namely: (i) the cosmological constant is obtained as an integration constant; (ii) the conservation of the energy-momentum tensor,  $\nabla_\mu T^\mu_{\phantom{\mu}\alpha}=0$, is a consequence of the gravitational field equation, rather than being assumed; and (iii) a conformally flat metric is not necessarily a vacuum solution.
More specifically, the field equations of CKG are given by 
\begin{equation}
H_{\alpha\mu\nu} = 8\pi G \, T_{\alpha\mu\nu},\label{eq_CKG}
\end{equation}
where $G$ is the gravitational constant, and throughout this work we use natural units, i.e. $G=c=1$. The tensors $H_{\alpha\mu\nu}$ and $T_{\alpha\mu\nu}$ are defined as
\begin{eqnarray}
H_{\alpha\mu\nu}&\equiv & \nabla_{\alpha}R_{\mu\nu}+\nabla_{\mu}R_{\nu\alpha}+\nabla_{\nu}R_{\alpha\mu}
	\nonumber \\
&&-\frac{1}{3}\left(g_{\mu\nu}\partial_{\alpha}+g_{\nu\alpha}\partial_{\mu}+g_{\alpha\mu}\partial_{\nu}\right)R,
	 \\
T_{\alpha\mu\nu}&\equiv & \nabla_{\alpha}T_{\mu\nu}+\nabla_{\mu}T_{\nu\alpha}+\nabla_{\nu}T_{\alpha\mu}
	\nonumber \\
&&-\frac{1}{6}\left(g_{\mu\nu}\partial_{\alpha}+g_{\nu\alpha}\partial_{\mu}+g_{\alpha\mu}\partial_{\nu}\right) T \,,
\end{eqnarray}
respectively, where $R_{\mu\nu}$ is the Ricci tensor and $T_{\mu\nu}$ is the energy-momentum tensor, with their respective traces $R$ and $T$. Note that $H_{\alpha\mu\nu}$ is totally symmetric in $\alpha$, $\mu$ and $\nu$ and satisfies $g^{\mu\nu}H_{\alpha\mu\nu}=0$ \cite{Harada}. Therefore, as $T_{\alpha\mu\nu}$ is also totally symmetric in $\alpha$, $\mu$ and $\nu$, and satisfies $g^{\mu\nu} T_{\alpha\mu\nu} =2 \nabla_\mu T^\mu_{\phantom{\mu}\alpha}$, consequently one obtains the conservation law $\nabla_\mu T^\mu_{\phantom{\mu}\alpha}=0$. Furthermore, solutions of GR are also solutions of CKG \cite{Harada}.

Taking into account a static and spherically symmetric metric of the form $ds^2=e^{a(r)}dt^2-e^{-a(r)}dr^2- r^2 \, d\Omega^2$, Harada also deduced the exact vacuum solution of the field equation $H_{\alpha\mu\nu} = 0 $, which is given by $e^{a(r)}=1 - \frac{2M}{r} - \frac{\Lambda}{3}r^2-\frac{\lambda}{5}r^4$ \cite{Harada}. Here,  the term $2M/r$ corresponds to the Schwarschild solution; the presence of $\Lambda r^2/3$ indicates a de Sitter term, where the cosmological constant $\Lambda$ is an integration constant, and the final term with $\lambda$ arises as the novel solution of the theory, which dominates at $r\rightarrow \infty$. If $\lambda =0 $, the solution reduces to the standard Schwarzschild-de Sitter solution of GR.  Furthermore, the most general spherically symmetric static vacuum solution of the theory was derived in \cite{Barnes:2023uru}. 
It was also recently shown that Eq. (\ref{eq_CKG}) is equivalent to Einstein's equation with an arbitrary conformal Killing tensor. This interesting realization reduces  the third order character with respect to the metric tensor of Eq. (\ref{eq_CKG}) to second order \cite{Mantica:2023stl}, and offers a simpler strategy of obtaining solutions of Harada's theory.

Recently, we further explored the theory by coupling CKG to nonlinear electrodynamics (NLED) and scalar fields, and found solutions of black holes and regular black holes \cite{Junior:2023ixh}. More specifically, by solving the field equations of CKG, we deduced forms for the NLED Lagrangian, the scalar field and the field potential, and analysed the regularity of the solutions through the Kretschmann scalar. We found generalizations of the Schwarschild--Reissner-Nordström--AdS solutions, and consequently further extended the class of (regular) black hole solutions found in the literature. As regular black holes are a topic currently of considerable interest in the GR and astrophysics communities,  it is interesting to note that a number of spacetimes which contains exact solutions  denoted as ``black bounces'' have been proposed in the literature \cite{Simpson:2018tsi}, that generalize and broaden the class of regular black holes beyond those usually considered \cite{Simpson:2019oft}.  The spacetime structure of a black bounce always features an event horizon. Within this horizon, there exists a bounce that leads from the interior region to another region of spacetime, which can be another part of our universe or a copy of our universe. 

The solution known as Simpson-Visser \cite{Simpson:2018tsi} has an event horizon and a bounce at $r=0$ within this horizon. The bounce connects the region where $r>0$ to another region with $r<0$, but passage in the opposite direction is not allowed. This solution also interpolates between a Schwarzschild solution, for the bounce parameter equal to zero, a regular black hole with a bounce at the origin, and a wormhole with a two-way timelike throat at $r=0$. In the case of the regular black hole with bounce, as within the event horizon the metric signature becomes $(-+--)$, the radial coordinate becomes a time coordinate and $t$ becomes spatial. Thus, this region is now dynamic and resembles a Kantowski-Sachs metric, and the function $\Sigma(r)=\sqrt{r^2+a^2}$ resembles a cosmological scale factor. Hence the name bounce for the origin $r=0$. This is not always true in general, as there exist other black bounces with two horizons, the event horizon and the Cauchy horizon, where in the region inside the latter, the metric signature changes to $(+---)$, and now the coordinates $t$ and $r$ are as in the exterior of the event horizon, akin to our usual universe \cite{Lobo:2020ffi}.

This class of solutions inspired research in the area and several solutions were found, such as specific solutions describing several physical situations of particular interest, including sources of black bounces \cite{Rodrigues2023}; a growing black-bounce, a wormhole to black-bounce transition, and the opposite black-bounce to wormhole transition were analysed in \cite{Simpson:2019cer}. In this reference, an Eddington-Finkelstein coordinate transformation is performed in which the constant mass parameter is now replaced by a dynamic mass that depends on the null temporal coordinate $w$. The resulting metric is a generalization of the Vaidya metric. One can also replace the constant parameter $a$ by a parameter that depends on the time coordinate $w$; electrically-charged wormhole and black hole solutions in Einstein-Maxwell-scalar theory, in which the scalar is a phantom field non-minimally coupled to the Maxwell field \cite{Huang:2019arj}; the stability of dynamic thin-shell black-bounce traversable wormholes were also explored \cite{Lobo:2020kxn}; the gravitational lensing was analysed in black bounce spacetimes that interpolate between regular black holes and traversable wormholes \cite{Nascimento:2020ime,Tsukamoto:2020bjm,Cheng:2021hoc,Tsukamoto:2021caq}; a plethora of novel geometries, more complex than before, with two or more horizons, with the possibility of an extremal case were found \cite{Lobo:2020ffi}, as well as charged black bounce solutions \cite{Franzin:2021vnj}; observational signatures were also explored \cite{Guerrero:2021ues,Jafarzade:2021umv,Yang:2021cvh,Bambhaniya:2021ugr,Ou:2021efv,Guo:2021wid,Wu:2022eiv,Tsukamoto:2022vkt,Zhang:2022nnj}; solutions in NLED and scalar fields were analysed \cite{Bronnikov:2022bud,Canate:2022gpy,Rodrigues2023,Pereira:2023lck};
and black bounce solutions were also studied in several modified theories of gravity \cite{Fitkevich:2022ior,Junior:2022zxo,Junior:2023qaq}, among other topics.

In this work, we build on the latter work, and explore black bounce solutions in CKG coupled to NLED and scalar fields. This work is outlined in the following manner: In Sec. \ref{sec2}, we briefly present the field equations of CKG coupled to NLED and scalar fields, and consider solutions solely described by magnetic charge. In Sec. \ref{sec5}, we analyse and generalize several black bounce geometries, and finally, in Sec.  \ref{sec:concl}, we summarize and discuss our results.

%\newpage

%%%%%%%%%%%%%%%%%%%%%%%%%%%%%%%%%%%%%%%%%%%%%%%%%
\section{CKG coupled to non-linear electrodynamics and scalar fields}\label{sec2}
%%%%%%%%%%%%%%%%%%%%%%%%%%%%%%%%%%%%%%%%%%%%%%%%%

As mentioned above, in this work, we couple CKG to NLED and scalar fields as matter sources applied to the energy-momentum tensor in the field equations~\eqref{eq_CKG}.  Thus, consider the that energy-momentum tensor is given by the following:
\begin{equation}
    T_{\alpha\mu\nu}=\overset{F}{T}_{\alpha\mu\nu}+\overset{\varphi}{T}_{\alpha\mu\nu}.\label{TEM}
\end{equation}
where the explicit contributions of the energy-momentum tensor described by the nonlinear electromagnetic field and the scalar field are given by:
\begin{eqnarray}
\overset{F}{T}_{\alpha\mu\nu}&\equiv & \nabla_{\alpha}\overset{F}{T}_{\mu\nu}+\nabla_{\mu}\overset{F}{T}_{\nu\alpha}+\nabla_{\nu}\overset{F}{T}_{\alpha\mu}
	\nonumber \\
&&-\frac{1}{6}\left(g_{\mu\nu}\partial_{\alpha}+g_{\nu\alpha}\partial_{\mu}+g_{\alpha\mu}\partial_{\nu}\right)\overset{F}{T},\\
\overset{\varphi}{T}_{\alpha\mu\nu}&\equiv &\nabla_{\alpha}\overset{\varphi}{T}_{\mu\nu}+\nabla_{\mu}\overset{\varphi}{T}_{\nu\alpha}+\nabla_{\nu}\overset{\varphi}{T}_{\alpha\mu}
	\nonumber \\
&&-\frac{1}{6}\left(g_{\mu\nu}\partial_{\alpha}+g_{\nu\alpha}\partial_{\mu}+g_{\alpha\mu}\partial_{\nu}\right)\overset{\varphi}{T} \,,
\end{eqnarray}
respectively, with
\begin{align}
    \overset{F}{T}{}_{\mu\nu}&=g_{\mu\nu}{\cal L}_{\rm NLED}(F)-{\cal L}_{F}F_{\mu\alpha}F_{\nu}^{\phantom{\nu}\alpha},\\
\overset{F}{T}&=4{\cal L}_{\rm NLED}(F)-4{\cal L}_{F}F,\\
    \overset{\varphi}{T}{}_{\mu\nu}&=2\,\epsilon\,\partial_{\mu}\varphi\partial_{\nu}\varphi-\epsilon 
g_{\mu\nu}\partial^{\sigma}\varphi\partial_{\sigma}\varphi+g_{\mu\nu}V(\varphi),\\
\overset{\varphi}{T}&=2\,\epsilon\,\partial^{\nu}\varphi\partial_{\nu}\varphi-\epsilon4\partial^{\sigma}\varphi\partial_{\sigma}\varphi+4V(\varphi).
\end{align}
\par
The nature of the scalar field $\varphi$ depends of the value of $\epsilon$, where $\epsilon=+1$ corresponds to a canonical scalar field, while $\epsilon=-1$ represents the phantom field; $V(\varphi)$ denotes the scalar potential,  ${\cal L}_{\rm NLED}(F)$ is the NLED Lagrangian density that depends of the electromagnetic scalar $F=\frac{1}{4}F^{\mu\nu}F_{\mu\nu}$,
%\begin{equation}
%F=\frac{1}{4}F^{\mu\nu}F_{\mu\nu},    \label{F}
%\end{equation}
 and the  Maxwell-Faraday antisymmetric tensor is defined by $F_{\mu \nu} = \partial_\mu A_\nu -\partial_\nu A_\mu$
%\begin{equation}
%   F_{\mu \nu} = \partial_\mu A_\nu -\partial_\nu A_\mu, 
%\end{equation}
 where ${A_\alpha}$ is the electromagnetic vector potential. 
We also present the following relevant expressions resulting from the influence of the gravitational field:
\begin{align}
\nabla_\mu ({\cal L}_F F^{\mu\nu})&=\frac{1}{\sqrt{-g} }\partial_\mu (\sqrt{-g} {\cal L}_F F^{\mu\nu})=0,\label{sol2}\\
     2\nabla_\nu\nabla^\mu \varphi&=-\frac{dV(\varphi)}{d\varphi}\,,
\end{align}
where ${\cal L}_F=\partial {\cal L} _{\rm NLED}(F)/\partial F$.

In the solutions obtained below, we will also consider the following useful consistency relationship
\begin{equation}
    {\cal L}_F-\frac{\partial {\cal L} _{\rm NLED}}{\partial r} \bigg(\frac{\partial F}{\partial r}\bigg)^{-1}=0.\label{RC}
\end{equation}

\par
Throughout this work we consider the following static and spherically symmetric metric:
\begin{equation}
ds^2=e^{a(r)}dt^2-e^{b(r)}dr^2-\Sigma^2(r) \, d\Omega^2,\label{m}
\end{equation}
where the metric functions $a(r), b(r)$ and $\Sigma(r)$ depend solely on the radial coordinate $r$, and $d\Omega^2\equiv d\theta^{2}+\sin^{2}\left(\theta\right)d\phi^{2}$.
%\begin{equation}
%d\Omega^2\equiv d\theta^{2}+\sin^{2}\left(\theta\right)d\phi^{2}.
%\end{equation}
The function $\Sigma(r)$ constrains the metric, since this quantity is determined by its shape~\eqref{m}. In the sections below, we will provide the corresponding metric for the solutions of black-bounces.

Here, we only consider solutions described by the magnetic charge, where the components for $F_{\mu\nu}$ and the electromagnetic scalar are $F_{23}=q \sin\theta$
and the electromagnetic scalar is given by $ F=\frac{q^2}{2 \Sigma^4(r)}$.
%\begin{equation}
%    F=\frac{q^2}{2 \Sigma^4(r)}.
%\end{equation}
The strategy that we adopt throughout this work consists essentially in the following: we consider a generalized metric that incorporates the term with $\lambda r^4$, that arises as the novel vacuum solution of the theory. Then, we integrate the gravitational field equations and determine the specific forms of the NLED Lagrangian and its derivative, by checking the consistency equation (\ref{RC}). Furthermore, we also check the regularity of the solutions by analysing the Kretschmann scalar.

%%%%%%%%%%%%%%%%%%%%%%%%%%%%%%%%%%%%%%%%%%%%%%%%%
\section{Black-Bounce solutions}\label{sec5}
%%%%%%%%%%%%%%%%%%%%%%%%%%%%%%%%%%%%%%%%%%%%%%%%%

As discussed in the introduction, the black-bounce is a spacetime that interpolates between a black hole, a regular black hole, and a wormhole. This type of solution has been studied in GR and in modified theories of gravity~\cite{Canate:2022gpy,Lima:2023jtl,Rodrigues2023,Lima:2023arg,Bronnikov:2023aya}.
 The matter source of these solutions was obtained by coupling the nonlinear electrodynamics and the scalar field with the potential. Thus, we follow
an analogous strategy and study the form of the nonlinear Lagrangian and the potential of the scalar field for the two examples that we discuss below.

In this section, we consider the contributions of NLED and a scalar field as the source of matter in the field equations \eqref{eq_CKG}. To derive these equations of motion, we use the metric \eqref{m} with the following expression 
\begin{equation}
    \Sigma(r)=\sqrt{L_0^2+r^2}\,,
\end{equation}
where the parameter $L_0 \in \Re$ is a regularization parameter and possesses the dimension of a length.

The metric \eqref{m} is now described as
\begin{equation}
   ds^2=e^{a(r)}dt^2-e^{b(r)}dr^2-\bigl(L_0^2 + r^2\bigr)\left[d\theta^{2}+\sin^{2}\left(\theta\right)d\phi^{2}\right],
   \label{m_BB}
\end{equation}
which is used to obtain black-bounce solutions in CKG.

%%%%%%%%%%%%%%%%%%%%%%%%%%%%%%%%%%%%%%%%%%%%%%%%%
%\subsection{First Simpson-Visser type black-bounce solution}\label{sec6A}
%%%%%%%%%%%%%%%%%%%%%%%%%%%%%%%%%%%%%%%%%%%%%%%%%

In developing our calculations for this model, we assume in the metric \eqref{m_BB} that the parameter $L_0 = q$ represents the magnetic charge. We also consider the symmetry with 
\begin{equation}
    a(r)=-b(r).\label{simet}
\end{equation}
Thus, by solving the equations of motion \eqref{eq_CKG} with the constraints mentioned above, i.e., using the metric \eqref{m_BB} described with the magnetic charge and the symmetry \eqref{simet}, we obtain the following quantities when solving the ``0, 0, 1'' and ``2, 1, 2'' components of the field equations:
\begin{widetext}
\begin{eqnarray}
{\cal L}_{\rm NLED}(r)&=& f_0-f_1 q^2 r^2-q^2 r^2\int e^{a(r)}\left\{ \frac{\left[8r^{3}-8re^{-a(r)}\left(q^{2}+r^{2}\right)\right]}{2\kappa^{2}q^{2}\left(q^{2}+r^{2}\right)^{4}}+\frac{2\epsilon\varphi'^{2}(r)\left[\left(q^{2}+r^{2}\right)a'(r)-2r\right]}{q^{2}\left(q^{2}+r^{2}\right)^{2}}+\frac{a^{(3)}(r)}{2\kappa^{2}q^{2}\left(q^{2}+r^{2}\right)} \right.	
	\nonumber\\
&&	
+\frac{a''(r)\left[3\left(q^{2}+r^{2}\right)a'(r)-2r\right]}{2\kappa^{2}q^{2}\left(q^{2}+r^{2}\right)^{2}}\left.+\frac{a'(r)\left\{ \left(q^{2}+r^{2}\right)a'(r)\left[\left(q^{2}+r^{2}\right)a'(r)-2r\right]+2\left(q-r\right)\left(q+r\right)\right\} }{2\kappa^{2}q^{2}\left(q^{2}+r^{2}\right)^{3}}\right\}\, dr
	\nonumber\\
&& + \int 	 e^{a(r)}\left\{ \frac{r^{2}a'(r)\left\{ \left(q^{2}+r^{2}\right)a'(r)\left[\left(q^{2}+r^{2}\right)a'(r)-2r\right]+2\left(q-r\right)\left(q+r\right)\right\} }{2\kappa^{2}\left(q^{2}+r^{2}\right)^{3}}+\frac{r^{2}a^{(3)}(r)}{2\kappa^{2}\left(q^{2}+r^{2}\right)} \right.	
	\nonumber\\	
	&&+\frac{4r\left(3q^{4}+3q^{2}r^{2}+r^{4}\right)-4r^{3}\left(q^{2}+r^{2}\right)}{\kappa^{2}\left(q^{2}+r^{2}\right)^{4}}+\frac{r^{2}a''(r)\left[3\left(q^{2}+r^{2}\right)a'(r)-2r\right]}{2\kappa^{2}\left(q^{2}+r^{2}\right)^{2}}
	\nonumber\\	
 &&\left.+2\epsilon\varphi'(r)\left(\frac{\varphi'(r)\left[\left(q^{2}+r^{2}\right)\left(q^{2}+2r^{2}\right)a'(r)+2q^{2}r\right]}{\left(q^{2}+r^{2}\right)^{2}}-2\varphi''(r)\right)-e^{-a(r)}V'(r)\right\} dr,	\label{L_BB} 
 \end{eqnarray}
 \begin{eqnarray}
{\cal L}_F(r) &=& f_1\left(q^2+r^2\right)^3+\left(q^2+r^2\right)^3\int e^{a(r)}\Bigg\{ \frac{\left[8r^{3}-8re^{-a(r)}\left(q^{2}+r^{2}\right)\right]}{2\kappa^{2}q^{2}\left(q^{2}+r^{2}\right)^{4}}+\frac{2\epsilon\varphi'^{2}(r)\left[\left(q^{2}+r^{2}\right)a'(r)-2r\right]}{q^{2}\left(q^{2}+r^{2}\right)^{2}} 
	\nonumber\\
&&
+\frac{a'(r)\left\{ \left(q^{2}+r^{2}\right)a'(r)\left[\left(q^{2}+r^{2}\right)a'(r)-2r\right]+2\left(q-r\right)\left(q+r\right)\right\} }{2\kappa^{2}q^{2}\left(q^{2}+r^{2}\right)^{3}} 
	\nonumber\\
&&+\frac{a^{(3)}(r)}{2\kappa^{2}q^{2}\left(q^{2}+r^{2}\right)}
+\frac{a''(r)\left[3\left(q^{2}+r^{2}\right)a'(r)-2r\right]}{2\kappa^{2}q^{2}\left(q^{2}+r^{2}\right)^{2}}
\Bigg\} 	dr.\label{LF_BB}
\end{eqnarray}
\end{widetext}

In the following, we consider the scalar field $\varphi$ according to the particular case of GR \cite{Rodrigues2023}, which is modeled as follows:
\begin{equation}
    \varphi(r)=\frac{\tan ^{-1}\left(\frac{r}{q}\right)}{\sqrt{\kappa ^2 (-\epsilon )}}.\label{phi_BB}
\end{equation}
Thus, substituting Eq. \eqref{phi_BB} in Eqs.~\eqref{L_BB} and~\eqref{LF_BB}, and consecutively   in condition \eqref{RC}, we arrive at:
\begin{equation}
    \frac{\left(q^{2}+r^{2}\right)}{2r}\left[-\frac{2e^{a(r)}a'(r)}{\kappa^{2}}-\frac{\left(q^{2}+r^{2}\right)^{2}V'(r)}{q^{2}}\right]=0\,,
    \label{rc_BB}
\end{equation}
which yields the following solution for the potential:
\begin{equation}
    V(r)= V_0 + \int -\frac{2 e^{a(r)} q^2 a'(r)}{\kappa ^2\left(q^2+r^2\right)^2 }dr.\label{V_BB}
\end{equation}

Starting from the metric \eqref{m_BB}, taking into account the symmetry \eqref{simet}  and consequently the results of the expressions for the Lagrangian and ${\cal L}_F(r)$, given by Eqs.~\eqref{L_BB} and \eqref{LF_BB}, respectively, and considering the scalar field \eqref{phi_BB}, which allows us to obtain $V(r)$, provided by Eq.~\eqref{V_BB}, we will use these results to find several solutions for black bounce models, namely, generalized Simpson-Visser type and Bardeen type solutions. We present these solutions in subsections \eqref{BB_SV} and \eqref{BB_Bardeen}.

%%%%%%%%%%%%%%%%%%%%%%%%%%%%%%%%%%%%%%%%%%%%%%%%%
\subsection{Simpson-Visser type solution}\label{BB_SV}
%%%%%%%%%%%%%%%%%%%%%%%%%%%%%%%%%%%%%%%%%%%%%%%%%

Consider the following  metric function
\begin{equation}
    e^{a(r)}=e^{-b(r)}= 1-\frac{2M}{\sqrt{q^{2}+r^{2}}}-\frac{\Lambda r^{2}}{3}-\frac{\lambda r^{4}}{5}.\label{a_BB}
\end{equation}

The formation of event horizons, taking into account the condition $g_{rr}^{-1}(r_H)=0$ (that, in both cases in this study, coincides with the Killing horizon $g_{tt}=0$), results from the condition:
\begin{equation}
     e^{a(r_{H})}=0,\label{rH}
\end{equation}
where the radius $r_{H}$ indicates the presence of the event horizon. For higher order polynomials, it is nontrivial to obtain analytic solutions of Eq. (\ref{rH}). Nevertheless, one can tackle the problem numerically, and classify the horizons, by imposing specific values of the parameters by consider the following  condition:
\begin{equation}
    \frac{d e^{a(r)}}{dr}\bigg|_{r=r_H}=0.
    \label{der_a}
\end{equation}
This relation provides the critical values of the parameters of the model, namely, $M$, $q$, $\Lambda$ or $\lambda$. Thus, here, we will analyze the presence of events horizons numerically by using simultaneously the conditions described by Eqs.~\eqref{rH} and~\eqref{der_a} to determine the radius of the horizon and the critical parameters from numerical solutions.

In the first solution, we consider $\Lambda < 0$, and choose the following values for the parameters of the model: $q = 0.5$, $\Lambda = -0.2$ and $\lambda=0$. Here, we assume that $\lambda=0$ as we will discuss later, this quantity causes the Kretschmann scalar to diverge. Thus, with these assumptions, we determine the value for the extreme mass from the simultaneous solution of Eqs. \eqref{rH} and~\eqref{der_a}, which is given by $M_c = 0.249$.

\begin{figure}[!h]
\centering
\includegraphics[scale=0.4]{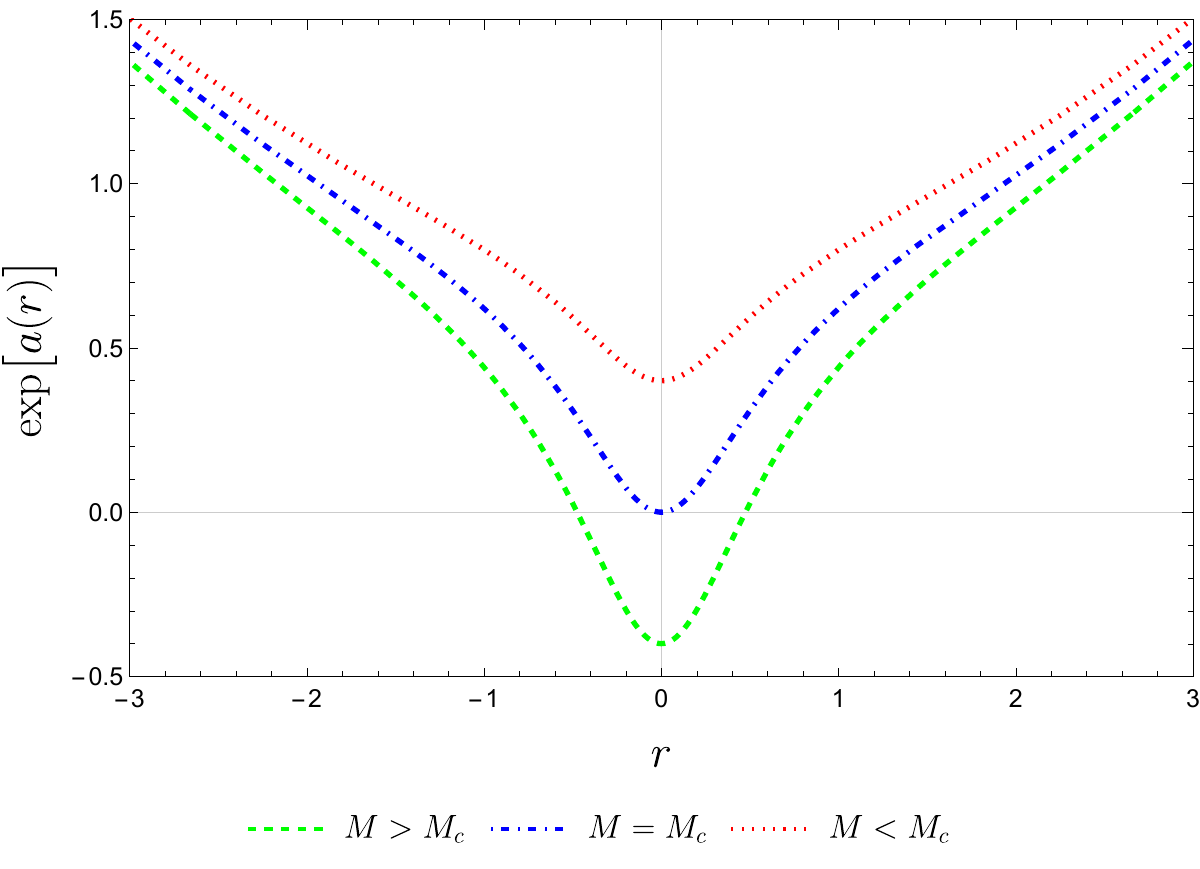}
\caption{The plot depicts $\exp{[a(r)]}$, described by Eq.~\eqref{a_BB}, for the values $\lambda=0$, $\Lambda =-0.2$ and $q=0.5$.  If $M > M_c$, there are two event horizons, one in each region $r>0$ and also in the region $r<0$. The metric function is negative within the event horizons, and there is a  bounce connecting these two regions. For $M=M_c$, we observe the presence of a double degenerate horizon and a bounce at $r=0$. When $M<M_c$, there are no horizons. In this case we have a wormhole.}
\label{figM1_BB}
\end{figure}

Figure~\ref{figM1_BB} depicts the behavior of the metric function~\eqref{a_BB} in terms of the radial coordinate $r$, with three mass scenarios: $M > M_c$, $M = M_c$, and $M < M_c$. When the mass exceeds the critical value, $M > M_c$, we observe the presence of an event horizon in the region where $r > 0$. Outside this horizon, the metric function $\exp{[a(r)]}$ is positive, while inside the event horizon, it is negative, and the signature of the metric in this latter case is  $(-,+,-,-)$. Furthermore, in this same scenario $(M > M_c)$, at $r=0$ we observe a bounce from the region $r > 0$ to the region $r < 0$, and conversely, due to the symmetry with respect to $r$, a bounce from the region $r < 0$ to the region $r > 0$ is also possible. The region in $r < 0$ is symmetrical and exhibits the same characteristics as the region $r > 0$, such as having only one event horizon. We note in this case that the metric function is always positive for $r \gg 1$ and $r \ll 1$.
In the case where the mass is equal to the critical mass, $M = M_c$, we observe a degenerate double horizon at $r=0$. However, at $r=0$, we have a bounce from the region $r > 0$ into the region $r < 0$, and as before, due to the symmetry in $r$ a bounce from the region $r < 0$ to the region $r > 0$ is also possible. In this case, the metric function is always positive for $r \gg 1$ and $r \ll 1$. However, when the mass is less than the critical mass, $M < M_c$, there is no formation of horizons, and the geometry we have is a wormhole with a traversable throat at the origin, i.e., $r=0$. Note that for these three cases described above, the metric function is always positive for $r \gg 1$ and $r \ll 1$.

\begin{figure}[!h]
\centering
\includegraphics[scale=0.40]{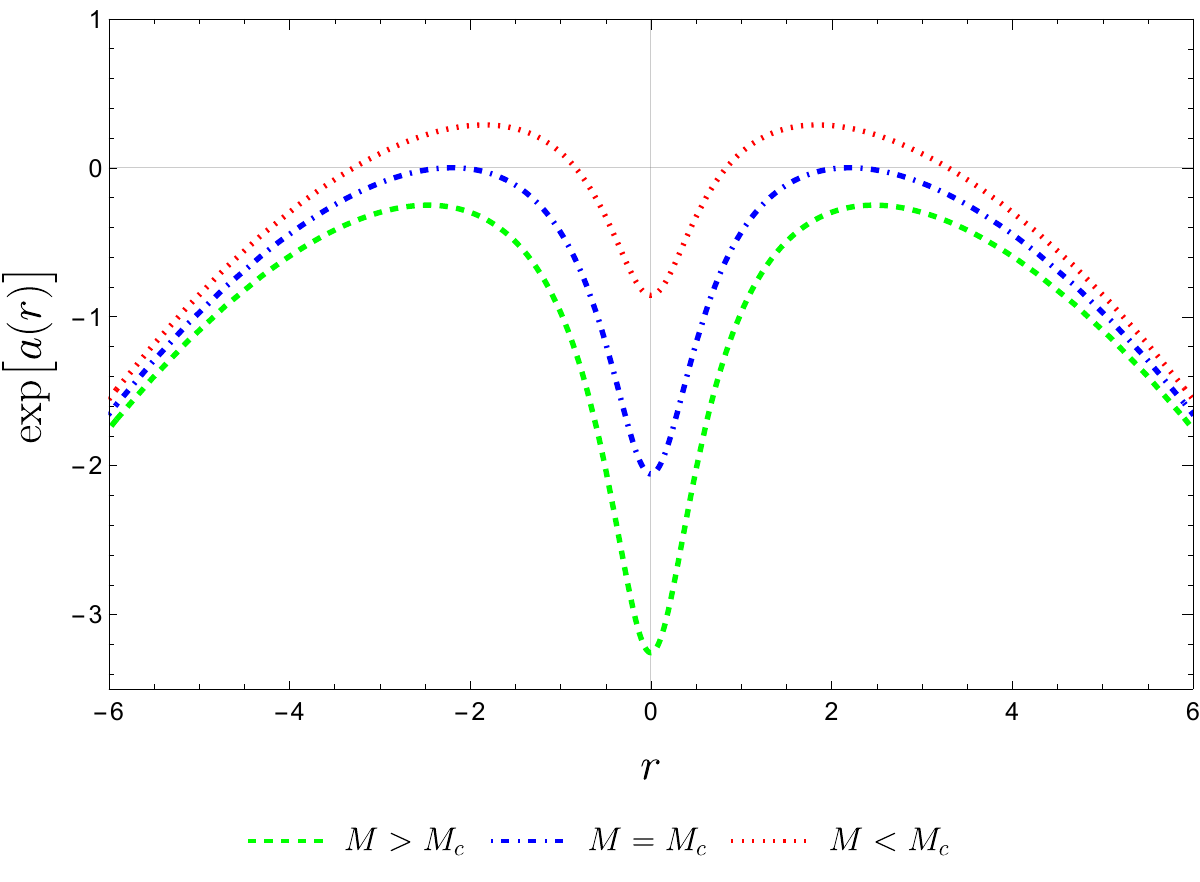}
\caption{The plot depicts $\exp{[a(r)]}$, given by Eq.~\eqref{a_BB},  with $\lambda=0$, $\Lambda =0.2$ and $q=0.5$.
When $M > M_c$, no horizons form, and the metric function is always negative, we have a wormhole. For $M=M_c$, we observe the presence of two degenerate horizons, one in the region $r>0$ and the other in the region $r<0$, along with a bounce at $r=0$. In the case where $M < M_c$, we find four horizons distributed in the regions $r>0$ and $r<0$: one event horizon and one cosmological horizon in each of these regions, in this case we have a bounce at $r=0$.
 }
\label{figM2_BB}
\end{figure}

In contrast, when analyzing the metric function~\eqref{a_BB} with $\Lambda > 0$, we verify the existence of a broader range of scenarios by varying the mass between $M > M_c$, $M = M_c$ and $M < M_c$. In this context, and as before, we assume certain values for the constants, for instance, considering $q = 0.5$, $\Lambda = 0.2$ and $\lambda=0$, the critical mass value is given by $M_c=0.764$. Lets begin our discussion by considering that the mass is less than the critical mass, $M < M_c$, where we observe more possibilities of horizons due to the positive cosmological constant $\Lambda > 0$. Thus, we note the formation of two different horizons in the region where $r > 0$: the first corresponds to the event horizon, and the second, the outermost one, is the cosmological horizon (we denote the radius of the cosmological horizon as $r_{\Lambda}$). Initially, we note that outside the cosmological horizon, the metric function $\exp{a(r)}$ is negative, and the metric signature is ($-,+,-,-$). Meanwhile, between  $r_H$ and $r_{\Lambda}$, the metric function $\exp{[a(r)]}$ is positive, and the metric signature now becomes ($+,-,-,-$). In turn, between $r=0$ and $r_H$, the function $\exp{[a(r)]}$ is again negative, implying the metric signature described again as ($-,+,-,-$). At $r=0$, we have a bounce from the region $r > 0$ into the region $r < 0$, and due to the symmetry with respect to $r$, a bounce from the region $r < 0$ to the region $r > 0$ is also possible. All these descriptions are present in the region $r < 0$ symmetrically to the region $r > 0$.
In the case where the mass is equal to the critical mass, $M = M_c$, we observe the existence of a degenerate horizon in the region where $r > 0$. At the center, at $r=0$, we observe a bounce from the region $r > 0$ to the region $r < 0$ and vice versa, due to the symmetry with respect to $r$, which also exhibits a horizon symmetric to that of the region $r > 0$.
When the mass exceeds the critical value, $M > M_c$, there is no formation of horizons, and the metric function remains negative in the regions where $r > 0$ and $r < 0$. Additionally, we observe at $r=0$ the presence of a traversable throat in both directions of the regions $r > 0$ and $r < 0$, configuring a wormhole.
We emphasize that the description of this spacetime is analogous to de Sitter's Schwarzschild model.  Finally, we note that the metric function is always negative at $r\gg1$ and $r\ll1$.  All this description is illustrated in Fig. \ref{figM2_BB}.
\par

\begin{figure}[!h]
\centering
\includegraphics[scale=0.4]{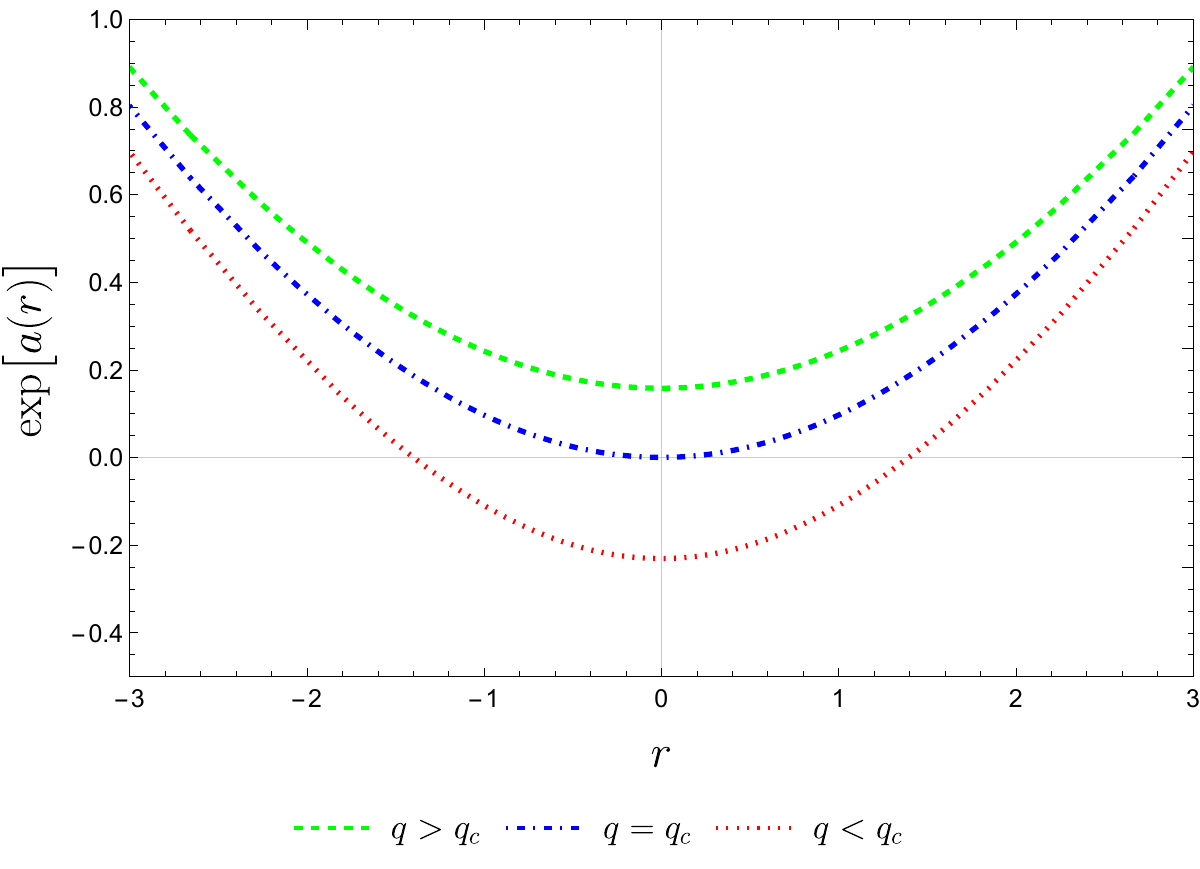}
\caption{The plot depicts $\exp{[a(r)]}$, described by Eq.~\eqref{a_BB}, for $\lambda=0$, $\Lambda =-0.2$ and $M=2.0$.
When $q>q_c$, we have a wormhole with a throat at $r=0$. When $q=q_c$,
we have a case similar to that in Fig. \ref{figM1_BB} for $M=M_c$.
In the case where $q<q_c$, the behavior corresponds to the case where $M > M_c$, as described in Fig. \ref{figM1_BB}.
 }  
\label{figq_BB}
\end{figure}

Now, considering the following values for the parameters: $\lambda=0$, $M=2.0$, and $\Lambda=-0.2$, the critical charge value is given by $q_c=4.0$. The behavior of $\exp{[a(r)]}$ is depicted in Fig.~\ref{figq_BB}, where the charge takes on the values $q > q_c$, $q=q_c$, and $q<q_c$.
Now, when $q>q_c$, there is no event horizon formation in the region where $r>0$, and consequently, there are no event horizons in the region $r<0$. This spacetime is described by a wormhole, and at $r=0$, we have a traversable  throat. Note that $\exp{[a(r)]}$ is always positive regardless of the value of $r$. In the configuration where $q=q_c$, we have a case similar to that in Fig. \ref{figM1_BB} for $M=M_c$. However, when the charge is less than the critical charge $q<q_c$, we find ourselves in a situation similar to the curve depicted in Fig.~\ref{figM1_BB}, which corresponds to the scenario where $M > M_c$. Finally, it is noteworthy that for values of $r$ much larger $r\gg 1$ and much smaller $r\ll 1$, the metric function remains always positive.

\begin{figure}[!h]
\centering
\includegraphics[scale=0.4]{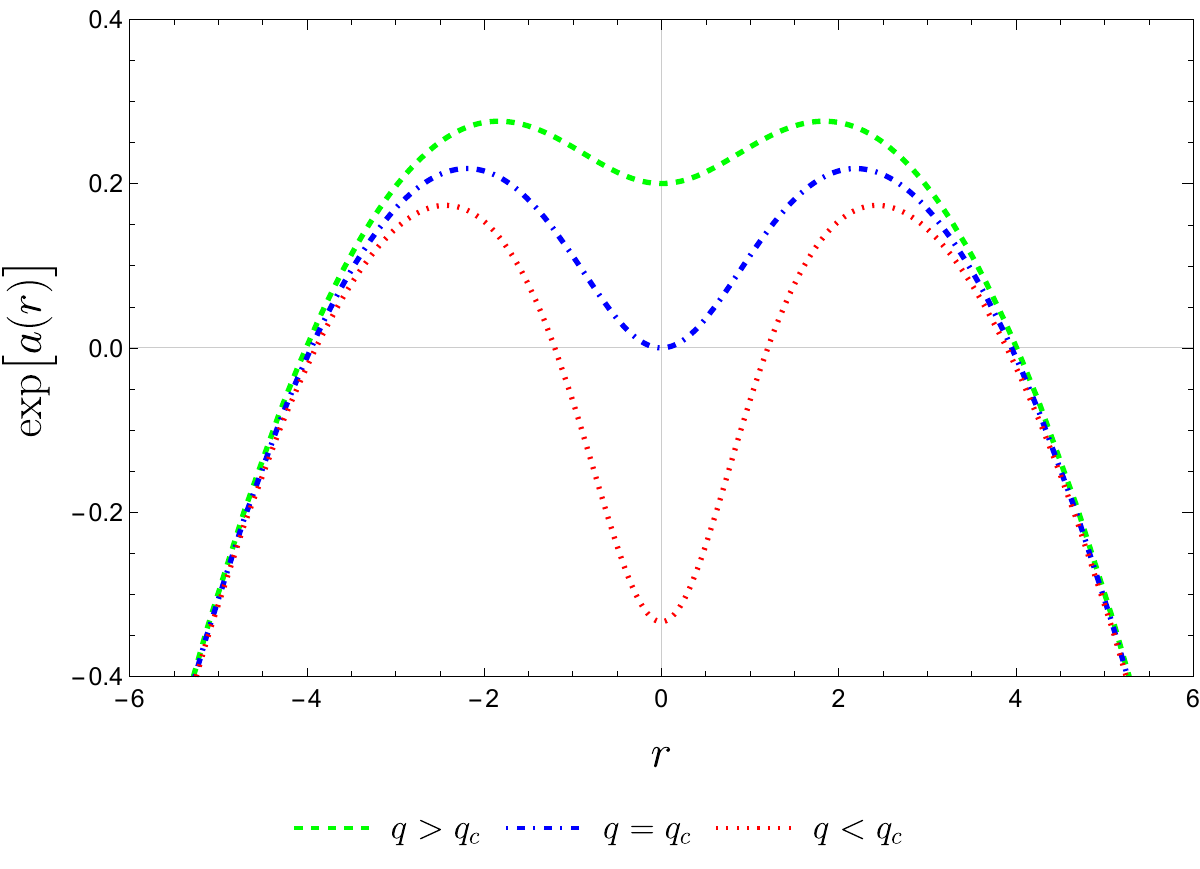}
\caption{The plot depicts $\exp{[a(r)]}$, given by Eq.~\eqref{a_BB}, with respect to the coordinate $r$, for the values $\lambda=0$, $\Lambda =0.12$ and $M=0.8$. When $q>q_c$, we observe the presence of two cosmological horizons, one in the region where $r>0$ and another in the region where $r<0$, while at $r=0$, in this case, we observe the presence of a traversable throat. In the case where $q=q_c$, we also find two cosmological horizons, one in each of the regions: $r>0$ and $r<0$. At $r=0$, there are degenerate horizons and a bounce. The scenario where $q<q_c$ is analogous to the case depicted in Fig. \ref{figM2_BB} with $M<M_c$.
}.
\label{figq2_BB}
\end{figure}

In addition to this, we have also developed numerical solutions with $\Lambda > 0$ for the critical charge. In this case, we consider the following constant values for the parameters: $\lambda=0$, $M=0.8$, and $\Lambda=0.12$. The critical charge for these conditions is $q_c=1.6$. The behavior of $\exp{[a(r)]}$ as a function of $r$ is shown in Fig.~\ref{figq2_BB}, where the charge takes the values $q > q_c$, $q=q_c$, and $q < q_c$.  When $q>q_c$, we observe the presence of only the cosmological horizon in the region where $r>0$. Outside of $r_{\Lambda}$, the metric function $\exp[a(r)]$ is negative, and the metric signature is $(-,+,-,-)$. Between the center $r=0$ and $r_{\Lambda}$, the metric function $\exp[a(r)]$ is positive. Once again at the origin, $r=0$, we observe the presence of a traversable throat, indicating a wormhole geometry. In the case with $q=q_c$, we have a cosmological horizon in the region where $r>0$, and at $r=0$, we have degenerate horizons and a bounce in both directions due to the symmetry of $r$.  Outside $r_{\Lambda}$ in the $r>0$ region, the metric function $\exp[a(r)]$ is negative, and the metric signature is ($-,+,-,-$). In turn, between $r=0$ and $r_{\Lambda}$, the metric function is positive. These aspects appear symmetrically in the region where $r<0$.
Finally, we have the case where $q<q_c$. This specific scenario is analogous to the case represented in Fig. \ref{figM2_BB} with $M<M_c$. Note that the metric function in the three cases described above is always negative for $r\gg1$ and $r\ll1$.

Now, using Eq. \eqref{a_BB}, and the scalar field $\varphi(r)$ given by (\ref{phi_BB}), we obtain $\cal{L}_{\rm NLED}$, ${\cal L}_F$ and $V(r)$, given by
\begin{eqnarray}
{\cal L}_{\rm NLED}(r) &=&  f_{0}+\frac{q^{2}}{15\kappa^{2}}\Bigg[
\frac{18M}{\left(q^{2}+r^{2}\right)^{5/2}}-\frac{5\Lambda+3\lambda r^{2}}{q^{2}+r^{2}}
		\nonumber \\
&& \hspace{-0.5cm} -15f_{1}\kappa^{2}r^{2}-9\lambda \Big]
-\frac{4\lambda q^{2}\textrm{ln}\left(q^{2}+r^{2}\right)}{5\kappa^{2}},
\label{L2_BB} 
\end{eqnarray}
\begin{eqnarray}
{\cal L}_F(r)=\left(q^2+r^2\right)^3 
\Big\{f_1+\Big[45 M+\left(q^2+r^2\right)^{3/2} \times
	\nonumber\\
\left(-5 \Lambda +15 \lambda  q^2+12 \lambda  r^2\right)\Big]
\Big/\left[15 \kappa ^2 \left(q^2+r^2\right)^{7/2}\right]\Big\},\label{LF2_BB}
\end{eqnarray}
\begin{eqnarray}
&V(r)= \frac{4q^{2}}{15\kappa^{2}}\left[\frac{3M}{\left(q^{2}+r^{2}\right)^{5/2}}+\frac{6\lambda q^{2}-5\Lambda}{2\left(q^{2}+r^{2}\right)}+3\lambda\textrm{ln}\left(q^{2}+r^{2}\right)\right]\,,
\nonumber\\
\label{V2_BB}
\end{eqnarray}
respectively.

Expressing $r(F)$ and $r(\cal{\varphi})$, we determine
\begin{eqnarray}
   {\cal L} _{\rm NLED} (F)  &=&	f_{0}-\frac{f_{1}q^{3}}{\sqrt{2 F}}+f_{1}q^{4}+\frac{12\sqrt[4]{2}MF^{5/4}}{5\kappa^{2}\sqrt{q}}
	\nonumber \\   
 && \hspace{-1.25cm}  -\frac{\sqrt{2F}\Lambda q}{3\kappa^{2}}+\frac{\lambda  q^2 \left( \sqrt{2F} q-4 \textrm{ln} \left(\frac{q}{\sqrt{F}}\right)-4+\textrm{ln}(4)\right)}{5 \kappa ^2},
 \nonumber \\
 \label{L3_BB} 
\end{eqnarray}   
   \begin{eqnarray}
V(\varphi) &=& \frac{4 M \cos ^6\left(\varphi  \sqrt{\kappa ^2 (-\epsilon )}\right) \sqrt{q^2 \sec ^2\left(\varphi  \sqrt{\kappa ^2 (-\epsilon )}\right)}}{5 \kappa ^2 q^4}
	\nonumber \\
	&& \hspace{-0.85cm} +\frac{4 \lambda  q^2 \cos ^2\left(\varphi  \sqrt{\kappa ^2 (-\epsilon )}\right)}{5 \kappa ^2}-\frac{2 \Lambda  \cos ^2\left(\varphi  \sqrt{\kappa ^2 (-\epsilon )}\right)}{3 \kappa ^2}
	\nonumber\\
&& \hspace{-0.85cm} +\frac{4 \lambda  q^2 \ln \left(q^2 \sec ^2\left(\varphi  \sqrt{\kappa ^2 (-\epsilon )}\right)\right)}{5 \kappa ^2}.\label{V3_BB}
\end{eqnarray}  

Note that considering $f_0=0$, $f_1=0$, $\lambda=0$ and $\Lambda=0$ in the Lagrangian \eqref{L2_BB}, we obtain the same expression as the Lagrangian of GR as found in Ref. \cite{Rodrigues2023}.
In expression \eqref{V3_BB} that the scalar field must necessarily be a phantom field, i.e. 
$\epsilon=-1$, as is transparent in Eq. (\ref{phi_BB}).

If we consider the limit of $r\rightarrow \infty$ in Eq.~\eqref{L3_BB}, which corresponds to the consideration of $F\sim0$, we obtain
\begin{eqnarray}
   {\cal L}_{\text{NLED}}(F) &=&f_0-\frac{f_{1}q^{3}}{\sqrt{2F}}+f_1 q^4-\frac{4\lambda q^{2}}{5\kappa^{2}}\ln\left(\frac{q}{\sqrt{F}}\right)\nonumber\\
  &&-\frac{4 \lambda  q^2}{5 \kappa ^2} +\frac{\lambda  q^2 \ln (4)}{5 \kappa ^2}.
\end{eqnarray}
Therefore, we see clearly that for this limit, this Lagrangian does not have linearity in $F$.

In GR, the Simpson-Visser solution can only be modeled by coupling NLED with a phantom scalar field (see Ref. \cite{Rodrigues2023} for more details). Here, since matter can take on a more general functional form, it is also possible to model a Simpson-Visser type solution using only NLED.

The Kretschmannn scalar is given by
\begin{widetext}
\begin{align}
    K=&\frac{4\Lambda\left\{ 6Mq^{2}\left(-q^{4}+8q^{2}r^{2}+r^{4}\right)+\sqrt{q^{2}+r^{2}}\left[\Lambda q^{8}+4\Lambda q^{6}r^{2}+8q^{4}r^{2}\left(2\Lambda r^{2}-3\right)+2q^{2}r^{4}\left(8\Lambda r^{2}-3\right)+6\Lambda r^{8}\right]\right\} }{9\left(q^{2}+r^{2}\right)^{9/2}}
    \nonumber\\
  & +\frac{4\left[3M^{2}\left(3q^{4}-4q^{2}r^{2}+4r^{4}\right)+8Mq^{2}\left(r^{2}-q^{2}\right)\sqrt{q^{2}+r^{2}}+3q^{4}\left(q^{2}+r^{2}\right)\right]}{\left(q^{2}+r^{2}\right)^{5}}
  \nonumber\\
 & +\frac{8\lambda\Lambda r^{2}\left(6q^{8}+24q^{6}r^{2}+52q^{4}r^{4}+46q^{2}r^{6}+15r^{8}\right)}{15\left(q^{2}+r^{2}\right)^{4}}+\frac{4\lambda r^{2}}{25\left(q^{2}+r^{2}\right)^{9/2}}\Bigg\{ 20M\left(-3q^{6}+6q^{4}r^{2}+8q^{2}r^{4}+3r^{6}\right)
 	\nonumber\\
& +r^{2}\sqrt{q^{2}+r^{2}}\left[36\lambda q^{8}+144\lambda q^{6}r^{2}+q^{4}\left(242\lambda r^{4}-60\right)+2q^{2}r^{2}\left(92\lambda r^{4}-15\right)+53\lambda r^{8}\right]\Bigg\}\,,
	\label{K_BB}
\end{align}
\end{widetext}
which is regular in the limit of $r\rightarrow0$.  Imposing $\lambda\rightarrow0$, the limit $r\rightarrow\infty$ is also regular.

The trace equation reads
\begin{equation}
   \kappa^2\Theta+R= 4 f_0 \kappa ^2-2 f_1 \kappa ^2 q^2 \left(q^2+3 r^2\right)-4 \Lambda -\frac{6 \lambda  q^2}{5}-6 \lambda  r^2,
   \label{sca_BB}
\end{equation}
which reduces to GR for $f_0=f_1=\lambda=0$.

\begin{figure}[!h]
\centering
\includegraphics[scale=0.47]{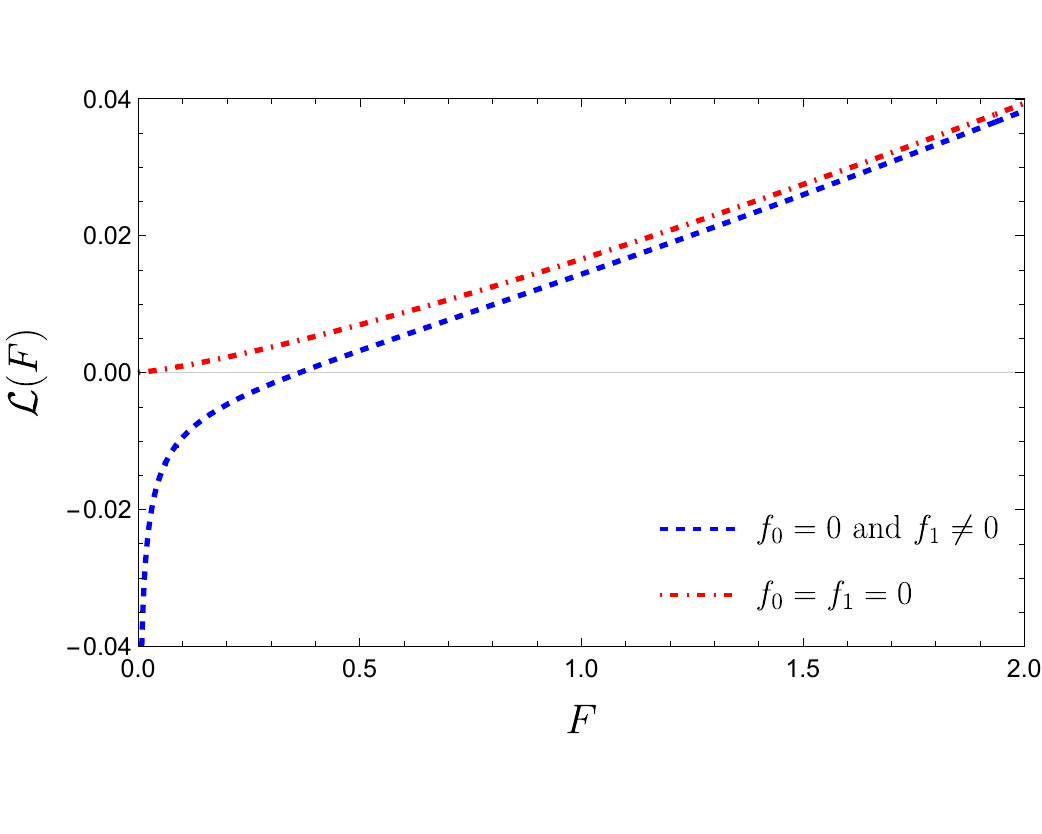}
\caption{The plot depicts ${\cal L}(F) $, given by Eq.~\eqref{L3_BB}, with respect to $F$. The blue dashed curve represents the behavior of ${\cal L}(F)$ versus $F$ with $f_0=0$ and $f_1=0.2$, while the red dotted-dashed curve illustrates the behavior of ${\cal L}(F)$ versus $F$ with $f_0=f_1=0$. The values of the constants used are  $\lambda=0$, $\Lambda=-0.2$,  $q=0.3$, $\kappa=8\pi$ and $M=2.0$. } 
\label{LxF_BB}
\end{figure}

\begin{figure}[!h]
\centering
\includegraphics[scale=0.47]{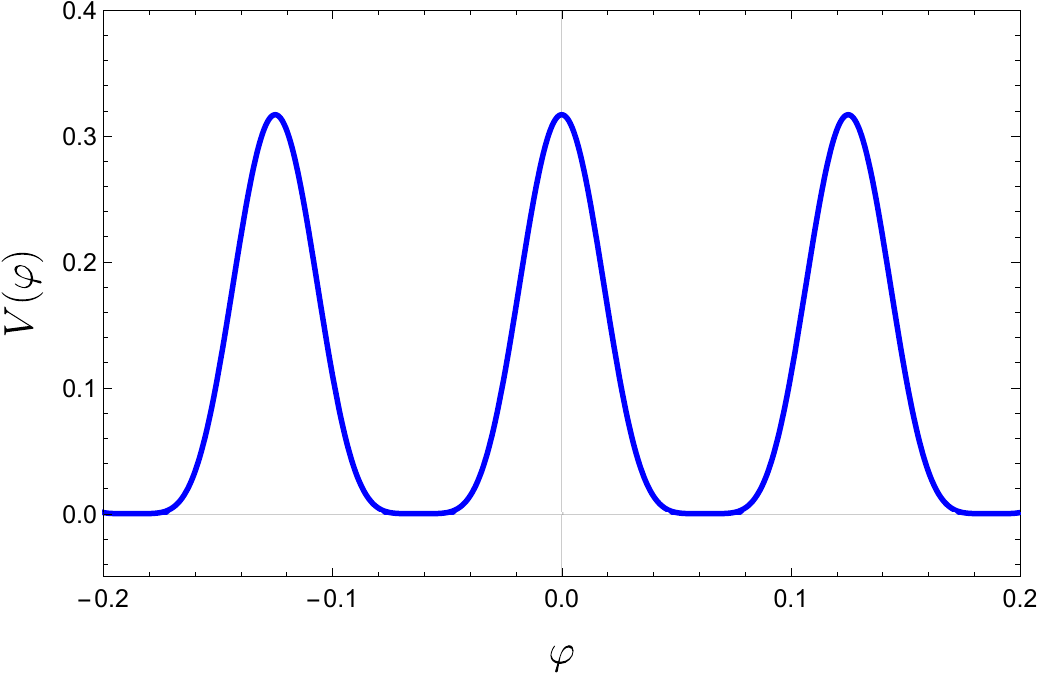}
\caption{The plot depicts $V(\varphi)$, described by Eq.~\eqref{V3_BB}, relative to $\varphi$, for the values $\lambda=0$, $\Lambda=-0.2$, $\kappa=8\pi$, $M=2.0$, $\epsilon =-1.0$ and $q=0.2$ .} 
\label{Vxphi_BB}
\end{figure}

Figure \eqref{LxF_BB} depicts the behavior of Eq.~\eqref{L3_BB} using the blue dashed curve for $f_0=0$ and $f_1=0.2$ and for the case similar to GR with the red dotted-dashed curve with the values $f_0=0$ and $f_1=0$.
In turn, we illustrate the behavior of Eq.~\eqref{V3_BB} with respect to $\varphi$ in Fig.~\ref{Vxphi_BB}.

\par

%%%%%%%%%%%%%%%%%%%%%%%%%%%%%%%%%%%%%%%%%%%%%%%%%
\subsection{Bardeen-type solution}\label{BB_Bardeen}
%%%%%%%%%%%%%%%%%%%%%%%%%%%%%%%%%%%%%%%%%%%%%%%%%

Consider now the following metric function:
\begin{equation}
    e^{a(r)}=e^{-b(r)}= 1-\frac{2 M r^2}{\left(q^2+r^2\right)^{3/2}}-\frac{\Lambda r^{2}}{3}-\frac{\lambda r^{4}}{5}.\label{a2_BB}
\end{equation}
In order to analyse the possibility of horizons we use Eqs.~\eqref{rH} and~\eqref{der_a} to determine the radius of the horizon and the critical parameters from numerical solutions. 

\begin{figure}[!h]
\centering
\includegraphics[scale=0.4]{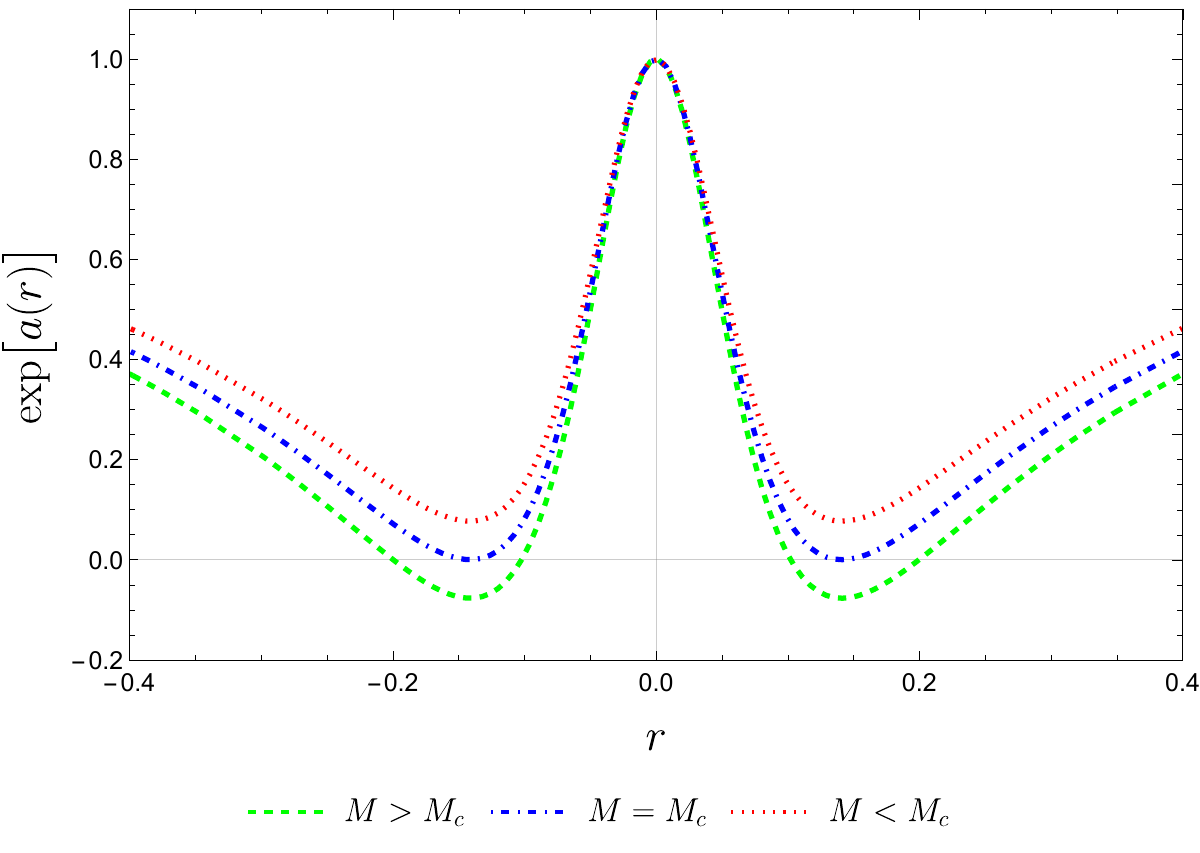}
\caption{The plot depicts $\exp{[a(r)]}$, described by Eq.~\eqref{a2_BB}, with respect to the coordinate $r$. We considered the values $\lambda=0$, $\Lambda =-0.2$ and $q=0.1$. When $M > M_c$, the Cauchy and event horizons form in the region where $r > 0$, and these horizons also occur symmetrically in the region where $r < 0$. In the case where $M = M_c$, degenerate horizons exist both in $r > 0$ and $r < 0$, with a throat at $r = 0$. When $M < M_c$, we observe that are no horizons, we have a wormhole with a throat at $r=0$}.
\label{figM2_BB2}
\end{figure}

Let us first look at $\Lambda < 0$ and choose the following parameter values for this model: $q = 0.1$, $\Lambda = -0.2$ and $\lambda=0$. We will see later that $\lambda \neq 0$ leads to a divergence of the Kretschmann scalar. With these assumptions, we determine the value of the critical mass from Eqs.~\eqref{rH} and ~\eqref{der_a} and obtain $M_c = 0.130$. The behavior of the metric function~\eqref{a2_BB} is shown in Fig.~\ref{figM2_BB2} for three mass scenarios: $M > M_c$, $M = M_c$, and $M < M_c$. When the mass surpasses the critical mass, $M > M_c$, we observe the formation of two horizons in the region where $r > 0$: the Cauchy horizon (the innermost one, denoted by the radius $r_C$) and the event horizon (the outermost one). In this same region, we note that outside the event horizon $r_H$, the metric function given by equation \eqref{a2_BB} is positive, and the metric signature is the standard $(+,-,-,-)$. Meanwhile, between the Cauchy horizon $r_C$ and $r_H$, the metric function is negative, and the metric signature is given by $(-,+,-,-)$. Between $r=0$ and $r_C$, the metric function is positive. At $r=0$, we have a throat, and the region at $r<0$ is symmetrical to $r>0$, presenting these two horizons, along with the other aspects discussed regarding the region where $r > 0$. In the case where $M=M_c$, we have degenerate horizons in $r > 0$ and $r < 0$, and now these regions are connected by a throat. In the case where $M < M_c$, we observe a scenario similar to that in Fig. \ref{figM1_BB} for $M < M_c$. In all these scenarios, we emphasize that $\exp[a(r)]$ is positive for $r \gg 1$ and also $r \ll 1$.

\begin{figure}[!h]
\centering
\includegraphics[scale=0.42]{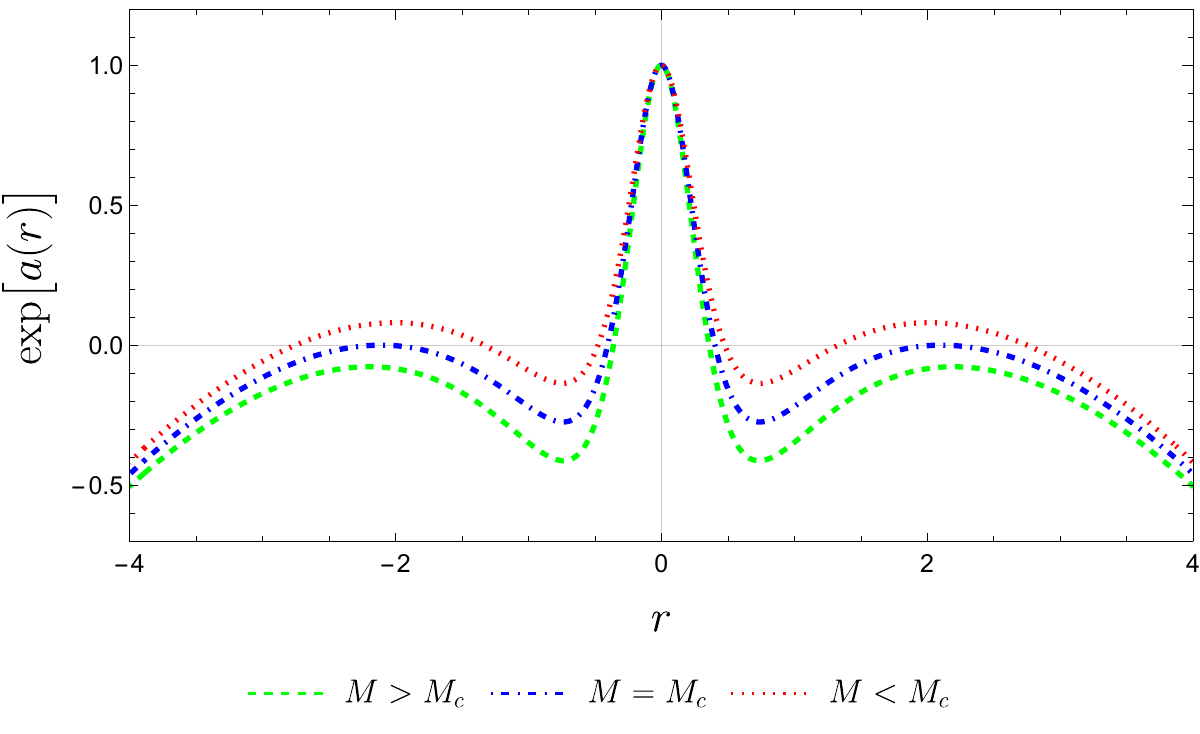}
\caption{The plot depicts $\exp{[a(r)]}$, given by Eq.~\eqref{a2_BB}, for the values $\lambda=0$, $\Lambda =0.2$ and $q=0.5$.
In the scenario where $M > M_c$, we observe the presence of an event horizon in the region where $r > 0$ and another symmetrically in the region where $r < 0$. At $r=0$, there is a  bounce from the region $r > 0$ to $r < 0$ and vice versa due to the symmetry of $r$.
In the case where $M=M_c$, the formation of two horizons occurs in both regions where $r > 0$ and $r < 0$. The innermost horizon is the Cauchy horizon, while the outermost horizon is the degenerate horizon. At $r=0$, these regions are connected by a bounce from the region $r > 0$ to the region $r < 0$, and a bounce from the region $r < 0$ to $r > 0$ is also possible due to the symmetry of $r$. In the last scenario, where $M < M_c$, we observe the formation of three horizons in the region where $r > 0$: the innermost is the Cauchy horizon, the intermediate is the event horizon, and the outermost is the cosmological horizon. In the region where $r < 0$, these three horizons are also observed symmetrically. At $r=0$, there is a bounce.
} 
\label{figM_BB2}
\end{figure}

Now, for $\Lambda > 0$, assuming the following values for the parameters: $q=0.5$, $\Lambda=0.2$, and $\lambda=0$,  this results in the value for the critical mass of $M_c=0.805$. 
In the scenario where $M > M_c$, the formation of an event horizon occurs in both the region where $r > 0$ and the region where $r < 0$. These regions are connected at $r=0$ by a  bounce from the region $r > 0$ to $r < 0$, and from region $r<0$ to $r>0$ is also possible, due to the symmetry with respect to $r$. In the region $r > 0$, the metric function is negative outside the event horizon, and the metric signature is $(-,+,-,-)$. Between $r=0$ and the event horizon, the metric function is positive. All these aspects are also evident in the region $r < 0$ for this case.
In the case where $M=M_c$, we observe the formation of two horizons in both the regions where $r > 0$ and $r < 0$. The innermost horizon is the Cauchy horizon ($r_C$), while the outermost horizon is the degenerate horizon, resulting from the union of the cosmological and event horizons in the case where $M<M_c$. At $r=0$, these regions are connected by a bounce in both regions of $r$, due to symmetry, i.e. from $r > 0$ to $r < 0$ and vice versa. Now, in the region where $r > 0$, outside the outermost horizon and between these two horizons, the metric function is negative, and the metric signature is $(-,+,-,-)$. Between $r=0$ and the radius $r_C$, the metric function is positive. These aspects are also evident in the region $r < 0$ for $M=M_c$. In the last scenario, where $M < M_c$, we observe the formation of three horizons: the innermost is the Cauchy horizon, the intermediate is the event horizon, and the outermost is the cosmological horizon. Outside the cosmological horizon $r_{\Lambda}$ and between $r_{C}$ and $r_H$, the metric function is negative, resulting in the metric signature $(-,+,-,-)$. Between the center of the black bounce and $r_{C}$, as well as between $r_H$ and $r_{\Lambda}$, the metric function is positive, maintaining the usual metric signature. At $r=0$, and again we have a bounce from the region $r > 0$ to the region $r < 0$ and due to the symmetry with respect to $r$ a bounce from the region $r < 0$ to the region $r > 0$ is also possible. All these descriptions made for $M < M_c$ in the region $r > 0$ are also present in the region $r < 0$. We also note that, in these three scenarios, the metric function is negative for $r \gg 1$ and also for $r \ll 1$. Fig. \ref{figM_BB2} illustrates all these situations.
\par

%%%%%% qc

\begin{figure}[!h]
\centering
\includegraphics[scale=0.4]{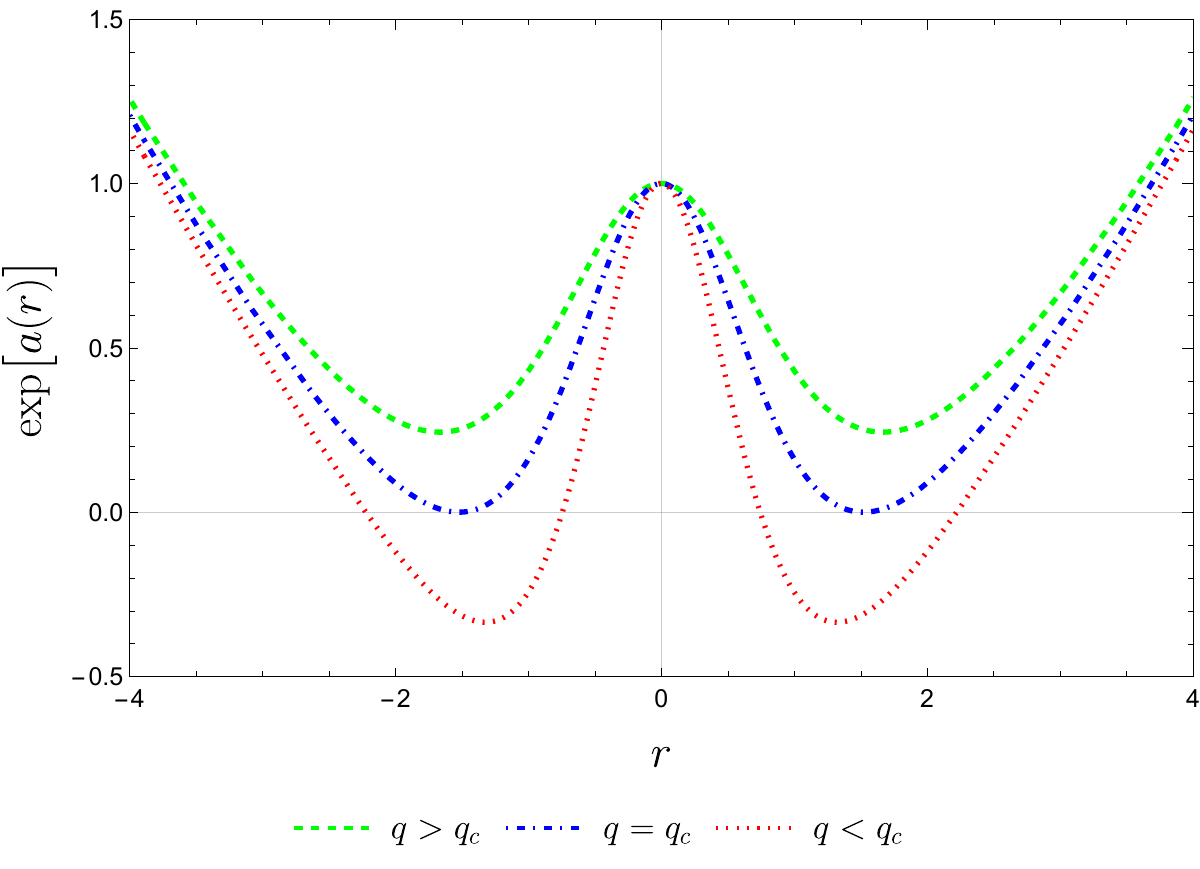}
\caption{The plot represents $\exp{[a(r)]}$, describe by Eq.~\eqref{a2_BB}, for $\lambda=0$, $\Lambda =-0.2$ and $M=2.0$. When the charge exceeds the critical charge, $q > q_c$, there are no horizons in either the $r>0$ or $r<0$ region. In the case where $q=q_c$, we observe the presence of two horizons, one in the region where $r>0$ and another symmetric horizon in $r<0$. The scenario where $q < q_c$ presents a Cauchy horizon and an event horizon in both regions: $r>0$ and $r<0$.} 
\label{figq_BB_Bardeen}
\end{figure}

Consider now the following parameter values: $\lambda=0$, $M=2.0$, and $\Lambda=-0.2$, which provides the value of  critical charge given by $q_c=1.300$. The behavior of $\exp[a(r)]$ is shown in Fig.~\ref{figq_BB_Bardeen}, for the ranges $q > q_c$, $q=q_c$ and $q <q_c$.
When the charge exceeds the critical charge, $q > q_c$, we observe a behavior similar to the curve described in Fig. \ref{figM2_BB2} for the case where $M <M_c$. Similarly, we can note the same resemblance in the behavior of the curve when $q=q_c$ compared to the curve for $M=M_c$ in Fig. \ref{figM2_BB2}. Furthermore, the scenario in which $q < q_c$ presents the same pattern of behavior as the case depicted in Fig. \ref{figM2_BB2}, but with $M > M_c$. Furthermore, we observe that, for very large values of $r$  and very small values of $r$, the metric function in this approach remains always positive.

\begin{figure}[!h]
\centering
\includegraphics[scale=0.4]{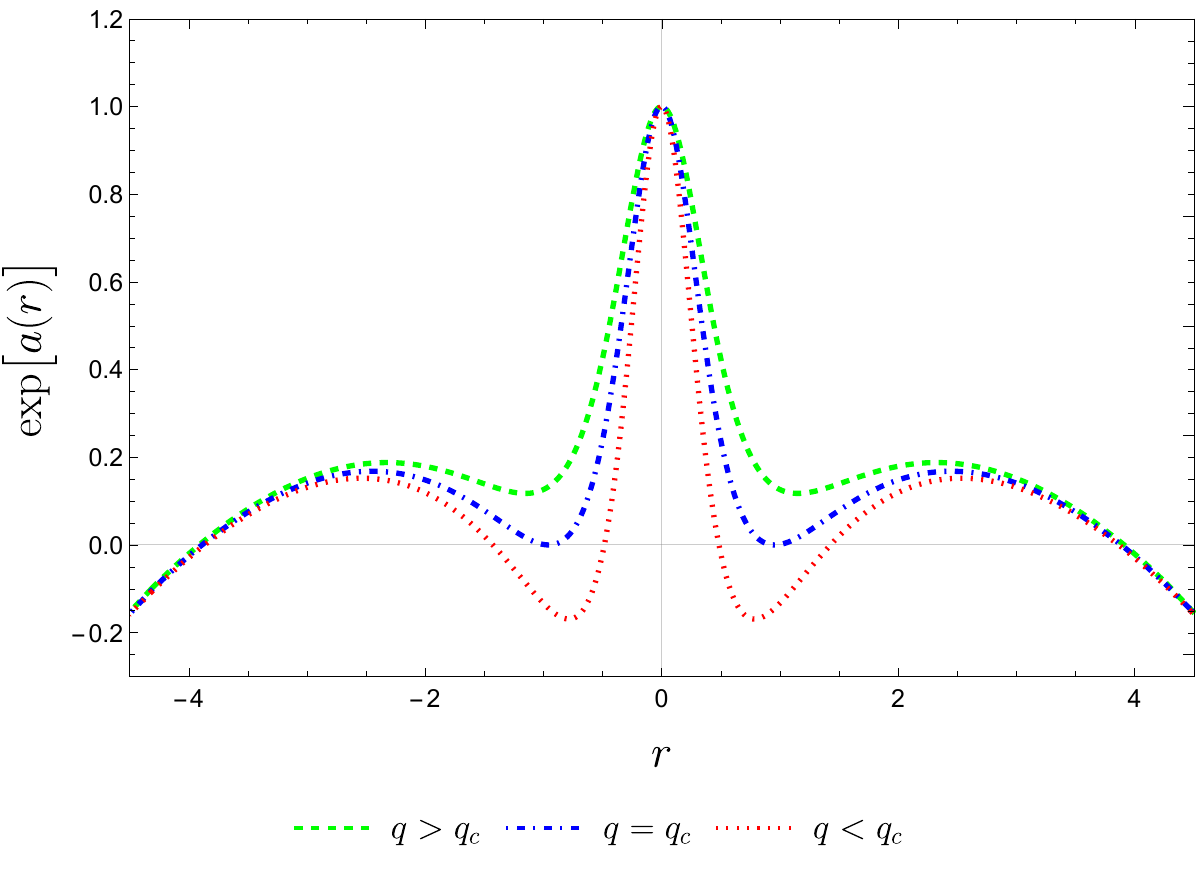}
\caption{The plot depicts $\exp{[a(r)]}$, given by Eq.~\eqref{a2_BB}, for $\lambda=0$, $\Lambda =0.12$ and $M=0.8$. In the scenario where $q > q_c$, we observe the presence of a cosmological horizon in the region where $r > 0$ and another symmetrically in the region where $r < 0$. %At $r=0$, there is a wormhole with a bidirectional throat similar to time.
In the case where $q=q_c$, the formation of two horizons occurs in both regions where $r > 0$ and $r < 0$. The innermost horizon is the Cauchy horizon, while the outermost horizon is the degenerate horizon. In the last scenario, where $q < q_c$, we observe the formation of three horizons in the region where $r > 0$ and also three symmetrical horizons in the region where $r<0$, with the innermost being the Cauchy horizon, the intermediate being the event horizon, and the outermost being the cosmological horizon.}
 
\label{figq2_BB_Bardeen}
\end{figure}

For the case $\Lambda > 0$, we consider the following parameter values: $\lambda = 0$, $M = 0.8$ and $\Lambda = 0.12$, so that the critical charge is given by $q_c = 0.637$. The behavior of $\exp[a(r)]$ is depicted in Fig.~\ref{figq2_BB_Bardeen}, where the charge assumes the values $q > q_c$, $q = q_c$ and $q < q_c$. When the charge exceeds the critical charge, $q > q_c$, we observe a behavior similar to the curve described in Fig. \ref{figq2_BB} for the case where $q > q_c$. In the case where $q=q_c$, we note the presence of a cosmological horizon and a degenerate horizon in both regions $r>0$ and $r<0$. At $r=0$, there is a bounce from the region $r > 0$ into the region $r < 0$, and vice-versa due to the symmetry in $r$. Outside the cosmological horizon, the metric function is negative, and the metric signature is ($-,+,-,-$). Meanwhile, between the bounce and the degenerate horizon and between the degenerate horizon and the cosmological horizon, the metric function in this case is positive. This is valid for the region $r<0$. In the scenario where $q < q_c$, it exhibits the same behavior as the case depicted in Fig. \ref{figM_BB2}, but with $M < M_c$. Finally, we note that for $r\gg1$ and $r\ll1$, the metric function is always negative. All these descriptions can be observed in Fig. \ref{figq2_BB_Bardeen}.
%%%%%%%%%%%%%%%%%%%%%%%%%%%%%%%%%%%%%%%%%%%%%%%%%%%%%%%%%%%%%%%%%%%%%%%%%%%%%%%%%%%%%%%%

Using Eq. \eqref{a_BB}, we obtain ${\cal L}_{\rm NLED}(r)$, ${\cal L}_F(r)$ and $V(r)$, which are expressed as follows
\begin{align} 
&{\cal L}_{\rm NLED}(r) = \frac{6M\left(16q^{4}+91q^{2}r^{2}\right)}{\left(105\kappa^{2}\right)\left(q^{2}+r^{2}\right)^{7/2}}-\frac{\left(84\lambda q^{2}\right)\ln\left(q^{2}+r^{2}\right)}{105\kappa^{2}}
\nonumber\\
&\phantom{{\cal L}_{\rm NLED}}-\frac{7q^{2}\left[15f_{1}\kappa^{2}r^{2}\left(q^{2}+r^{2}\right)+5\Lambda+9\lambda q^{2}+12\lambda r^{2}\right]}{\left(105\kappa^{2}\right)\left(q^{2}+r^{2}\right)}
\nonumber\\
&\phantom{{\cal L}_{\rm NLED}}+\frac{105f_{0}\kappa^{2}}{105\kappa^{2}},\label{L_BB2} 
\end{align}
\begin{eqnarray} 
{\cal L}_F(r)&=&\left(q^2+r^2\right)^3 
\Big\{f_1+\Big[M\left(195r^{2}-30q^{2}\right)+\big(-5\Lambda
	\nonumber\\
&&\hspace{-1.5cm}+15\lambda q^{2}+12 \lambda  r^2\big)
\left(q^{2}+r^{2}\right)^{5/2}\Big]\Big/\left[15 \kappa ^2 \left(q^2+r^2\right)^{9/2}\right]\Big\} ,\label{LF_BB2} \nonumber \\
\end{eqnarray} 
\begin{eqnarray} 
V(r)&=&-\frac{2q^{2}}{105\kappa^{2}}\Bigg[\frac{6M\left(8q^{2}-7r^{2}\right)}{\left(q^{2}+r^{2}\right)^{7/2}}-\frac{7\left(6\lambda q^{2}-5\Lambda\right)}{q^{2}+r^{2}}
\nonumber\\
&&-42\lambda\ln\left(q^{2}+r^{2}\right)\Big] \,,
\label{V2_BB20} 
\end{eqnarray}
respectively.

Expressing $r(F)$ and $r(\cal{\varphi})$, we determine
\begin{eqnarray}
   {\cal L} _{\rm NLED} (F)  &= &	f_{0}-\frac{f_{1}q^{3}}{\sqrt{2F}}+f_{1}q^{4}-\frac{60\ 2^{3/4}F^{7/4}M\sqrt{q}}{7\kappa^{2}}
	\nonumber \\   
 && +\frac{52\sqrt[4]{2}F^{5/4}M}{5\kappa^{2}\sqrt{q}}+\frac{\sqrt{2F}\lambda q^{3}}{5\kappa^{2}}-\frac{4\lambda q^{2}}{5\kappa^{2}}
 \nonumber \\
&& \hspace{-1cm} -\frac{4\lambda q^{2}\ln\left(\frac{q}{\sqrt{F}}\right)}{5\kappa^{2}}-\frac{\sqrt{2F}\Lambda q}{3\kappa^{2}}+\frac{\lambda q^{2}\ln(4)}{5\kappa^{2}}.
 \label{L2_BB2} 
\end{eqnarray}  
And
\begin{eqnarray}
V(\varphi) &=& -\frac{32M\cos^{8}\left(\text{\ensuremath{\varphi}}\sqrt{\kappa^{2}(-\epsilon)}\right)\sqrt{q^{2}\sec^{2}\left(\text{\ensuremath{\varphi}}\sqrt{\kappa^{2}(-\epsilon)}\right)}}{35\kappa^{2}q^{4}}\nonumber\\
&& \hspace{-1.2cm}  -\frac{2\Lambda\cos^{2}\left(\text{\ensuremath{\varphi}}\sqrt{\kappa^{2}(-\epsilon)}\right)}{3\kappa^{2}}
+\frac{4\lambda q^{2}\ln\left[q^{2}\sec^{2}\left(\text{\ensuremath{\varphi}}\sqrt{\kappa^{2}(-\epsilon)}\right)\right]}{5\kappa^{2}}
	\nonumber \\
&& +\frac{4M\sin^{2}\left(\text{\ensuremath{\varphi}}\sqrt{\kappa^{2}(-\epsilon)}\right)\cos^{6}\left(\text{\ensuremath{\varphi}}\sqrt{\kappa^{2}(-\epsilon)}\right)}{5\kappa^{2}q^{4}}\times\nonumber\\
&& \hspace{-1.2cm}  \times\sqrt{q^{2}\sec^{2}\left(\text{\ensuremath{\varphi}}\sqrt{\kappa^{2}(-\epsilon)}\right)}+\frac{4\lambda q^{2}\cos^{2}\left(\text{\ensuremath{\varphi}}\sqrt{\kappa^{2}(-\epsilon)}\right)}{5\kappa^{2}}
	.\label{V2_BB2}
\end{eqnarray} 

Considering again $f_0=0$, $f_1=0$, $\lambda=0$ and $\Lambda=0$ in the Lagrangian of equation~\eqref{L2_BB}, we obtain the same expression as the Lagrangian of GR, found in Ref.~\cite{Rodrigues2023}.

If we take the limit of $r\rightarrow \infty$ in Eq.~\eqref{L2_BB2} again, we now obtain
\begin{eqnarray}
   {\cal L}_{\text{NLED}}(F) &=&f_0+f_1 q^4-\frac{f_{1}q^{3}}{\sqrt{2F}}-\frac{4\lambda q^{2}}{5\kappa^{2}}\ln\left(\frac{q}{\sqrt{F}}\right)\nonumber\\
   &&-\frac{4 \lambda  q^2}{5 \kappa ^2}+\frac{\lambda q^{2}\ln(4)}{5\kappa^{2}}.
\end{eqnarray}
Therefore, we observe once again that, in this result above, this Lagrangian is not linear in $F$.

\begin{figure}[!h]
\centering
\includegraphics[scale=0.45]{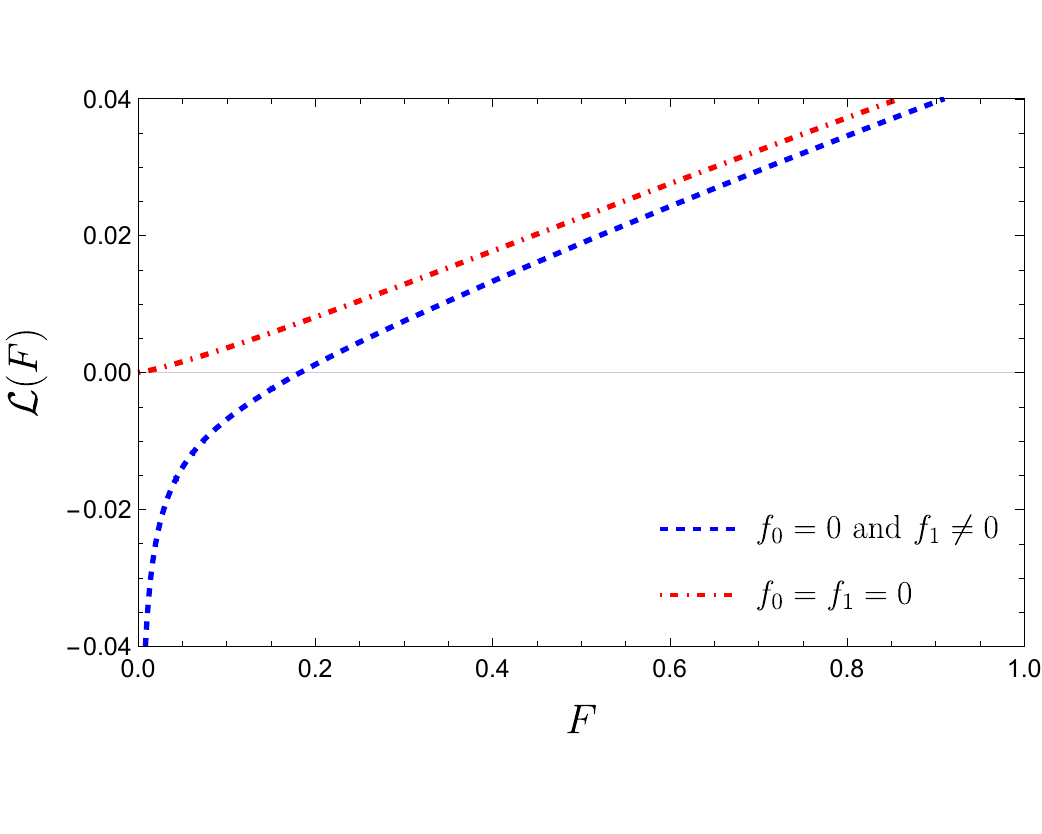}
\caption{Graphical representation of the expression ${\cal L}(F) $, described by Eq.~\eqref{L2_BB2}, with respect to the $F$. The blue dashed curve represents the behavior of ${\cal L}(F)$ versus $F$ with $f_0=0$ and $f_1=0.2$, while the red dotted-dashed curve illustrates the behavior of ${\cal L}(F)$ versus $F$ with $f_0=f_1=0$. The values of the constants used are  $\lambda=0$, $\Lambda=-0.2$,   $q=0.3$, $\kappa=8\pi$ and $M=2.0$ } 
\label{LxF_BB2}
\end{figure}

\begin{figure}[!h]
\centering
\includegraphics[scale=0.45]{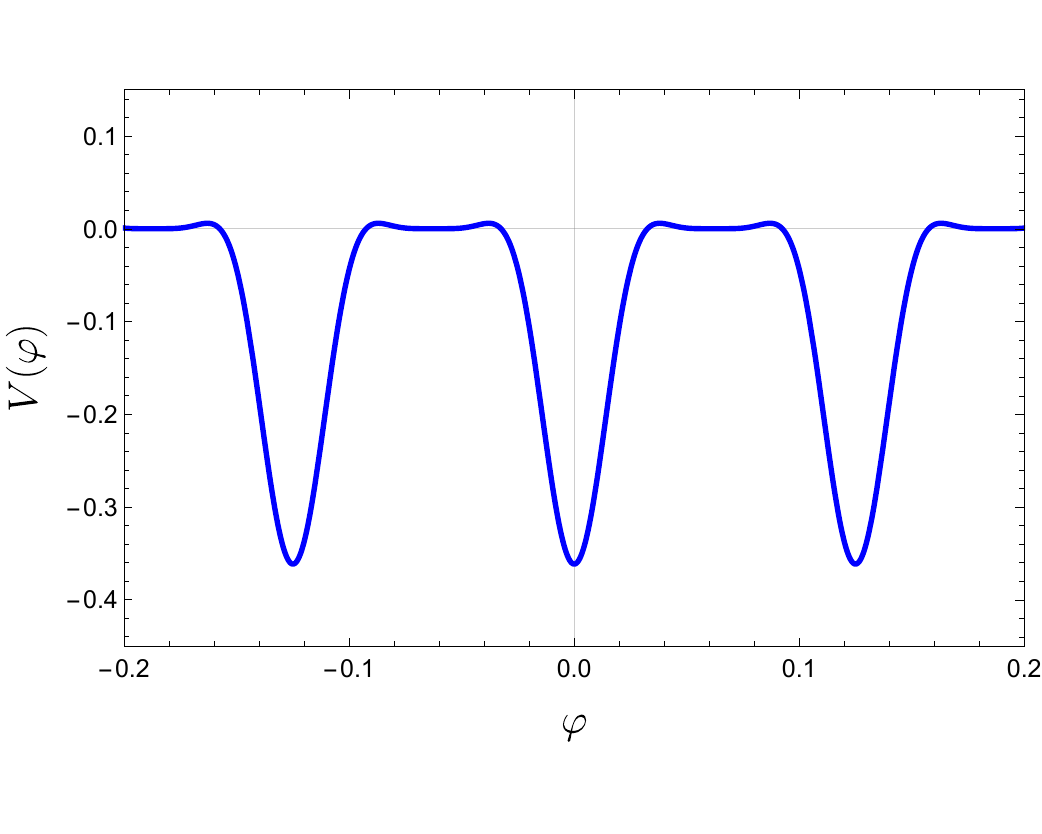}
\caption{Graphical representation of  $V(\varphi)$, describe by expression~\eqref{V2_BB2}, as a function of $\varphi$, for the specific values $\lambda=0$, $\Lambda=-0.2$, $\kappa=8\pi$, $M=2.0$, $\epsilon =-1.0$ and $q=0.2$ .} 
\label{Vxphi_BB2}
\end{figure}

Figure \ref{LxF_BB2} shows the behavior of Eq.~\eqref{L2_BB2} with the dashed blue curve for the values $f_0=0$ and $f_1=0.2$. The dotted-dashed  red curve represents a similar case to GR with the values $f_0=0$ and $f_1=0$. We also illustrate the behavior of Eq. $\eqref{V2_BB2}$ with respect to $\varphi$ in Fig.~\ref{Vxphi_BB2}.

In this case, the Kretschmann scalar takes the following form:
\begin{widetext}
\begin{align}
    &K=\frac{4}{225 \left(q^2+r^2\right)^4}\Big\{q^8 \left(5 \Lambda +18 \lambda  r^2\right)^2+4 q^6 \left(18 \lambda  r^3+5 \Lambda  r\right)^2+3r^{8}\left(50\Lambda^{2}+159\lambda^{2}r^{4}+150\lambda\Lambda r^{2}\right)
    \nonumber\\
    & + q^{4}\left[18\lambda r^{4}\left(121\lambda r^{4}-30\right)+400\Lambda^{2}r^{4}+120\Lambda r^{2}\left(13\lambda r^{4}-5\right)+675\right]
    \nonumber\\
    & +2q^{2}r^{4}\left\{ r^{2}\left[200\Lambda^{2}+9\lambda\left(92\lambda r^{4}-15\right)+690\lambda\Lambda r^{2}\right]-75\Lambda\right\} \Big\}+\frac{4 M^2 \left(4 q^8-44 q^6 r^2+169 q^4 r^4-68 q^2 r^6+12 r^8\right)}{\left(q^2+r^2\right)^7}+
    \nonumber\\
    &\frac{8 M \left[2 q^8 \left(5 \Lambda +18 \lambda  r^2\right)-7 q^6 \left(18 \lambda  r^4+5 \Lambda  r^2\right)-2 q^4 r^2 \left(114 \lambda  r^4-5 \Lambda  r^2+60\right)+q^2 r^4 \left(-84 \lambda  r^4-5 \Lambda  r^2+60\right)+18 \lambda  r^{10}\right]}{15 \left(q^2+r^2\right)^{11/2}}.
	\label{K_BB2}
\end{align}
\end{widetext}
In the limit of $r\rightarrow0$, the Kretschmann scalar is regular, and for $r\rightarrow\infty$, the regularity is imposed if $\lambda\rightarrow0$.

The trace equation reads
\begin{equation}
   \kappa^2\Theta+R= 4 f_0 \kappa ^2-2 f_1 \kappa ^2 q^2 \left(q^2+3 r^2\right)-4 \Lambda -\frac{6 \lambda  q^2}{5}-6 \lambda  r^2,
   \label{sca_BB2}
\end{equation}
which also reduces to GR for $f_0=f_1=\lambda=0$.

\par

%%%%%%%%%%%%%%%%%%%%%%%%%%%%%%%%%%%%%%%%%%%%%%%%%
\section{Summary and Conclusion}\label{sec:concl}
%%%%%%%%%%%%%%%%%%%%%%%%%%%%%%%%%%%%%%%%%%%%%%%%%

In this paper, we explored black bounce geometries within a novel gravitational framework recently developed known as conformal Killing gravity (CKG). It is noteworthy that this theory simultaneously satisfies three fundamental theoretical criteria for gravitational theories, which is not achieved by previous gravitational theories, including GR. We investigated black bounce solutions by coupling the CKG field equations with NLED and a scalar field. In particular, we studied two generalizations of black bounce solutions, namely, the Simpson-Visser type and the Bardeen type solutions.
More specifically, we developed black-bounce solutions by considering NLED and a scalar field as the source of the matter, by using the symmetry given by $a(r)=-b(r)$ metric functions. We found the ${\cal L}_{\rm NLED }(r)$ and ${\cal L}_F(r)$ functions from the ``0,0,1" and ``2,1,1" components of the gravitational field equations.  To progress in our calculations, we considered a particular scalar field from GR, according to Eq.~\eqref{phi_BB}, and used a Simpson-Visser and Bardeen type metric function, which differs from GR by the additional terms with $\lambda$ and $\Lambda$.

In both GR and conformal Killing gravity, a necessary
condition for modeling black bounces is the coupling of NLED and the scalar field with the potential. In this context, in GR, it has been shown that the solutions of black bounces in the form of equations \eqref{m}-\eqref{simet} are exact solutions of the equations of motion with a source of Nonlinear Electrodynamics (NLED) matter coupled with a scalar field \cite{Bronnikov:2021uta}. While it is possible that black bounce solutions can be modeled with the coupling of other types of matter, no models have been developed in the literature with matter fields other than those we have discussed. Furthermore, the representation of the matter field may not be unique. 

In the first Simpson-Visser model, we simultaneously solve equations~\eqref{rH} and~\eqref{der_a} for $\Lambda<0$, when $M > M_c$, we find two horizons. For $M = M_c$, a bounce occurs at $r=0$. If $M < M_c$, no horizon is identified. In the case of $\Lambda > 0$, no horizon formation occurs for $M > M_c$, while for $M = M_c$, degenerate horizons are observed. Lastly, when $M < M_c$, four horizons are found, including two cosmological horizons.
For $\Lambda < 0$, in the case of critical charge, we observe that there is no horizon for $q > q_c$. For $q < q_c$, we find two horizons. In the scenario where $\Lambda > 0$, we identify two cosmological horizons, in all scenarios of $q > q_c$, $q = q_c$, and $q < q_c$.

By analyzing the Kretschmann scalar according to Eq.~\eqref{K_BB} for this case, we see that it remains regular when we take the limit of $r\rightarrow 0$. In the limit of $r\rightarrow \infty$, on the other hand, we verify that the regularity in spacetime only occurs for the specific case of $\lambda=0$. Moreover, it is possible to recover GR if we examine the trace described by Eq.~\eqref{sca_BB}, it is possible to recover GR if we set $f_0=f_1=\lambda=0$.
We have also obtained the expressions for ${\cal L}(F)$ and $V(\varphi)$ as shown in Eqs.~\eqref{L3_BB} and \eqref{V3_BB}. The behavior of the first quantity is shown in Fig.~\ref{LxF_BB} with two curves, one of them for the case of $f_0=0$ and $f_1=0$, in which we recover the model described by GR. In Fig.~\ref{Vxphi_BB}, on the other hand, we verify the periodic nature of the potential.

In the second model, we adopted the Bardeen-type metric function and numerically determined the solutions for the horizons. For $\Lambda < 0$, in determining the critical mass, we observed the formation of four horizons when $M > M_c$, with the innermost ones being the Cauchy horizons. For $M = M_c$, degenerate horizons are observed. Finally, if $M < M_c$, the geometry corresponds to that of a wormhole.
For $\Lambda > 0$, when $M > M_c$, we observe the presence of two horizons. When $M = M_c$, we find four horizons, with the innermost one being the Cauchy horizon. In the scenario where $M < M_c$, we detect six horizons, with the outermost ones being the cosmological horizons.

In the scenario where $\Lambda < 0$, at the critical charge, if $q > q_c$, no horizon forms. For $q = q_c$, we have two degenerate horizons. When $q < q_c$, we observe the presence of four horizons, noting that the innermost ones are the Cauchy horizons.
When $\Lambda > 0$, we also note that in all scenarios ($q > q_c$, $q = q_c$, $q < q_c$), there are cosmological horizons. It is noteworthy that for $q < q_c$, we observe a total of six horizons.

By analyzing the Kretschmann scalar of the Bardeen-type model according to Eq.~\eqref{K_BB2}, we find that it remains regular as we approach the limit $r\rightarrow 0$. However, we only find the regularity of spacetime in the limit of $r\rightarrow \infty$ when $\lambda=0$. If we also examine the trace described in Eq.~\eqref{sca_BB2}, we find GR again when $f_0=f_1=\lambda=0$.
Finally, we obtained the expressions for ${\cal L}(F)$ and $V(\varphi)$ as described in Eqs. \eqref{L2_BB2} and \eqref{V2_BB2}., respectively. Again, we plotted the behavior of these quantities, where the first illustrates ${\cal L}(F)$ in Fig.~\ref{LxF_BB2} with two curves, one for the case $f_0=0$ and $f_1=0$, in which we recover the model described by GR. The behavior of the potential was represented in Fig.~\ref{Vxphi_BB2}.

To ensure accuracy, we calculated the numerical value of lambda $\lambda$ based on observations from the `Event Horizon Telescope' (EHT) collaboration of Sgr A*, using the photon sphere radius as 2.9 MKeck, the mass as $M\rightarrow MKeck=5.827\times10^9  m$, the charge as $q=0.5 MKeck$, the distance is $DKeck=2.4543\times10^{20}  m$, the distance from the observer is $r_0=10^{-10}$ DKeck  and $\Lambda=1.3\times10^{-52} m^{-2}$~\cite{Do:2019txf}. In this analysis, we verified that the obtained value  is given by $\lambda=5.902\times 10^{-43} m^{-4}$. 
Thus, based on these parameters, we have calculated the shadow radius of the black hole for the CKG model, and taking into account this specific value of $\lambda$, we verify that the shadow radius of the black hole of CKG agrees with the result estimated with EHT for Sgr. A*.

Although the stability issues are an important aspect
of these solutions, this analysis lies outside the scope of the present paper. However, it is interesting to note that one may explore the stability of our solutions in an analogous manner as outlined in Ref. \cite{Nashed:2021pah} for the selfgravitating spherically symmetric solutions found in $f(T)$ torsion gravity \cite{Nashed:2021pah,Kofinas:2015hla}. Here, the authors used a perturbative approach by considering small deviations from GR and found charged black hole solutions. The stability of the motion around the obtained solutions, was explored by analyzing the geodesic deviation, and the unstable regimes in the parameter space was found. Additionally, a detailed thermodynamic analysis was carried out by examining the temperature, entropy, heat capacity and Gibb’s free energy, and the analysis showed that $f(T)$ modifications of GR improve the thermodynamic stability, which is not the case in other classes of modified gravity. We aim to perform a similar analysis in future work.

As a follow-up to this work, we plan to investigate several other approaches within CKG, which include comprehensive studies of black hole thermodynamics, stability of solutions through perturbation analysis, detailed studies of black hole shadows, and gravitational lensing analysis. These are several of the areas we plan to address in our future work to deepen our understanding and further contribute to the development of this generalised new alternative approach to GR.

%\vspace{1cm}

%%%%%%%%%%%%%%%%%%%%%%%%%%%%%%%%%%%%%%%%%%%%%%%%%
\acknowledgments{
%{\bf Acknowledgement}:  
FSNL acknowledges support from the Funda\c{c}\~{a}o para a Ci\^{e}ncia e a Tecnologia (FCT) Scientific Employment Stimulus contract with reference CEECINST/00032/2018, and funding through the research grants UIDB/04434/2020, UIDP/04434/2020 and PTDC/FIS-AST/0054/2021.
MER thanks CNPq for partial financial support.  This study was supported in part by the Coordenção de Aperfeioamento de Pessoal de Nível Superior - Brazil (CAPES) - Financial Code 001.  }
%%%%%%%%%%%%%%%%%%%%%%%%%%%%%%%%%%%%%%%%%%%%%%%%%

%%%%%%%%%%%%%%%%%%%%%%%%%%%%%%%%%%%%%%%%%%%%%%%%%

%


\begin{thebibliography}{99}
%%%%%%%%%%%%%%%%%%%%%%%%%%%%%%%%%%%%%%%%%%%%%%%%%


%\cite{Harada}
\bibitem{Harada}
J.~Harada,
``Gravity at cosmological distances: Explaining the accelerating expansion without dark energy,''
Phys. Rev. D \textbf{108}, (2023) no.4, 044031
%doi:10.1103/PhysRevD.108.044031
[arXiv:2308.02115 [gr-qc]].
%2 citations counted in INSPIRE as of 04 Sep 2023

%\cite{Mantica:2023stl}
\bibitem{Mantica:2023stl}
C.~A.~Mantica and L.~G.~Molinari,
``A note on Harada's Conformal Killing gravity,''
[arXiv:2308.06803 [gr-qc]].
%1 citations counted in INSPIRE as of 04 Sep 2023

%\cite{Barnes:2023uru}
\bibitem{Barnes:2023uru}
A.~Barnes,
``Vacuum Static Spherically Symmetric Spacetimes in Harada's Theory,''
[arXiv:2309.05336 [gr-qc]].

%\cite{Junior:2023ixh}
\bibitem{Junior:2023ixh}
J.~T.~S.~S.~Junior, F.~S.~N.~Lobo and M.~E.~Rodrigues,
``(Regular) Black holes in conformal Killing gravity coupled to nonlinear electrodynamics and scalar fields,''
[arXiv:2310.19508 [gr-qc]].


%\cite{Simpson:2018tsi}
\bibitem{Simpson:2018tsi}
A.~Simpson and M.~Visser,
``Black-bounce to traversable wormhole,''
JCAP \textbf{02}, 042 (2019)
%doi:10.1088/1475-7516/2019/02/042
[arXiv:1812.07114 [gr-qc]].
%174 citations counted in INSPIRE as of 23 Sep 2023

%\cite{Simpson:2019oft}
\bibitem{Simpson:2019oft}
A.~Simpson,
``Traversable Wormholes, Regular Black Holes, and Black-Bounces,''
[arXiv:2104.14055 [gr-qc]].
%5 citations counted in INSPIRE as of 23 Sep 2023


\bibitem{Rodrigues2023}
M.~E.~Rodrigues and M.~V.~d.~S.~Silva,
``Source of black bounces in general relativity,''
Phys. Rev. D \textbf{107}  no.4, 044064 (2023)
%doi:10.1103/PhysRevD.107.044064
[arXiv:2302.10772 [gr-qc]].



%\cite{Simpson:2019cer}
\bibitem{Simpson:2019cer}
A.~Simpson, P.~Martin-Moruno and M.~Visser,
``Vaidya spacetimes, black-bounces, and traversable wormholes,''
Class. Quant. Grav. \textbf{36}, no.14, 145007 (2019)
%doi:10.1088/1361-6382/ab28a5
[arXiv:1902.04232 [gr-qc]].
%70 citations counted in INSPIRE as of 23 Sep 2023


%\cite{Huang:2019arj}
\bibitem{Huang:2019arj}
H.~Huang and J.~Yang,
``Charged Ellis Wormhole and Black Bounce,''
Phys. Rev. D \textbf{100}, no.12, 124063 (2019)
%doi:10.1103/PhysRevD.100.124063
[arXiv:1909.04603 [gr-qc]].
%35 citations counted in INSPIRE as of 23 Sep 2023

%\cite{Lobo:2020kxn}
\bibitem{Lobo:2020kxn}
F.~S.~N.~Lobo, A.~Simpson and M.~Visser,
``Dynamic thin-shell black-bounce traversable wormholes,''
Phys. Rev. D \textbf{101}, no.12, 124035 (2020)
%doi:10.1103/PhysRevD.101.124035
[arXiv:2003.09419 [gr-qc]].
%57 citations counted in INSPIRE as of 23 Sep 2023

%\cite{Nascimento:2020ime}
\bibitem{Nascimento:2020ime}
J.~R.~Nascimento, A.~Y.~Petrov, P.~J.~Porfirio and A.~R.~Soares,
``Gravitational lensing in black-bounce spacetimes,''
Phys. Rev. D \textbf{102}, no.4, 044021 (2020)
%doi:10.1103/PhysRevD.102.044021
[arXiv:2005.13096 [gr-qc]].
%44 citations counted in INSPIRE as of 23 Sep 2023

\cite{Tsukamoto:2020bjm}
\bibitem{Tsukamoto:2020bjm}
N.~Tsukamoto,
``Gravitational lensing in the Simpson-Visser black-bounce spacetime in a strong deflection limit,''
Phys. Rev. D \textbf{103}, no.2, 024033 (2021)
%doi:10.1103/PhysRevD.103.024033
[arXiv:2011.03932 [gr-qc]].
%54 citations counted in INSPIRE as of 23 Sep 2023

%\cite{Cheng:2021hoc}
\bibitem{Cheng:2021hoc}
X.~T.~Cheng and Y.~Xie,
``Probing a black-bounce, traversable wormhole with weak deflection gravitational lensing,''
Phys. Rev. D \textbf{103}, no.6, 064040 (2021).
%doi:10.1103/PhysRevD.103.064040
%41 citations counted in INSPIRE as of 23 Sep 2023

%\cite{Tsukamoto:2021caq}
\bibitem{Tsukamoto:2021caq}
N.~Tsukamoto,
``Gravitational lensing by two photon spheres in a black-bounce spacetime in strong deflection limits,''
Phys. Rev. D \textbf{104}, no.6, 064022 (2021)
%doi:10.1103/PhysRevD.104.064022
[arXiv:2105.14336 [gr-qc]].
%37 citations counted in INSPIRE as of 23 Sep 2023

%\cite{Lobo:2020ffi}
\bibitem{Lobo:2020ffi}
F.~S.~N.~Lobo, M.~E.~Rodrigues, M.~V.~de Sousa Silva, A.~Simpson and M.~Visser,
``Novel black-bounce spacetimes: wormholes, regularity, energy conditions, and causal structure,''
Phys. Rev. D \textbf{103}, no.8, 084052 (2021)
%doi:10.1103/PhysRevD.103.084052
[arXiv:2009.12057 [gr-qc]].
%95 citations counted in INSPIRE as of 23 Sep 2023

%\cite{Franzin:2021vnj}
\bibitem{Franzin:2021vnj}
E.~Franzin, S.~Liberati, J.~Mazza, A.~Simpson and M.~Visser,
``Charged black-bounce spacetimes,''
JCAP \textbf{07}, 036 (2021)
%doi:10.1088/1475-7516/2021/07/036
[arXiv:2104.11376 [gr-qc]].
%69 citations counted in INSPIRE as of 23 Sep 2023


%\cite{Guerrero:2021ues}
\bibitem{Guerrero:2021ues}
M.~Guerrero, G.~J.~Olmo, D.~Rubiera-Garcia and D.~S.~C.~G\'omez,
``Shadows and optical appearance of black bounces illuminated by a thin accretion disk,''
JCAP \textbf{08}, 036 (2021)
%doi:10.1088/1475-7516/2021/08/036
[arXiv:2105.15073 [gr-qc]].
%60 citations counted in INSPIRE as of 23 Sep 2023

%\cite{Jafarzade:2021umv}
\bibitem{Jafarzade:2021umv}
K.~Jafarzade, M.~Kord Zangeneh and F.~S.~N.~Lobo,
``Observational optical constraints of regular black holes,''
Annals Phys. \textbf{446}, 169126 (2022)
%doi:10.1016/j.aop.2022.169126
[arXiv:2106.13893 [gr-qc]].
%11 citations counted in INSPIRE as of 23 Sep 2023

%\cite{Yang:2021cvh}
\bibitem{Yang:2021cvh}
Y.~Yang, D.~Liu, Z.~Xu, Y.~Xing, S.~Wu and Z.~W.~Long,
``Echoes of novel black-bounce spacetimes,''
Phys. Rev. D \textbf{104}, no.10, 104021 (2021)
%doi:10.1103/PhysRevD.104.104021
[arXiv:2107.06554 [gr-qc]].
%28 citations counted in INSPIRE as of 23 Sep 2023

%\cite{Bambhaniya:2021ugr}
\bibitem{Bambhaniya:2021ugr}
P.~Bambhaniya, S.~K, K.~Jusufi and P.~S.~Joshi,
``Thin accretion disk in the Simpson-Visser black-bounce and wormhole spacetimes,''
Phys. Rev. D \textbf{105}, no.2, 023021 (2022)
%doi:10.1103/PhysRevD.105.023021
[arXiv:2109.15054 [gr-qc]].
%29 citations counted in INSPIRE as of 23 Sep 2023

%\cite{Ou:2021efv}
\bibitem{Ou:2021efv}
M.~Y.~Ou, M.~Y.~Lai and H.~Huang,
``Echoes from asymmetric wormholes and black bounce,''
Eur. Phys. J. C \textbf{82}, no.5, 452 (2022)
%doi:10.1140/epjc/s10052-022-10421-x
[arXiv:2111.13890 [gr-qc]].
%15 citations counted in INSPIRE as of 23 Sep 2023

%\cite{Guo:2021wid}
\bibitem{Guo:2021wid}
Y.~Guo and Y.~G.~Miao,
``Charged black-bounce spacetimes: Photon rings, shadows and observational appearances,''
Nucl. Phys. B \textbf{983}, 115938 (2022)
%doi:10.1016/j.nuclphysb.2022.115938
[arXiv:2112.01747 [gr-qc]].
%19 citations counted in INSPIRE as of 23 Sep 2023

%\cite{Wu:2022eiv}
\bibitem{Wu:2022eiv}
S.~R.~Wu, B.~Q.~Wang, D.~Liu and Z.~W.~Long,
``Echoes of charged black-bounce spacetimes,''
Eur. Phys. J. C \textbf{82}, no.11, 998 (2022)
%doi:10.1140/epjc/s10052-022-10938-1
[arXiv:2201.08415 [gr-qc]].
%5 citations counted in INSPIRE as of 23 Sep 2023


%\cite{Tsukamoto:2022vkt}
\bibitem{Tsukamoto:2022vkt}
N.~Tsukamoto,
``Retrolensing by two photon spheres of a black-bounce spacetime,''
Phys. Rev. D \textbf{105}, no.8, 084036 (2022)
%doi:10.1103/PhysRevD.105.084036
[arXiv:2202.09641 [gr-qc]].
%12 citations counted in INSPIRE as of 23 Sep 2023

%\cite{Zhang:2022nnj}
\bibitem{Zhang:2022nnj}
J.~Zhang and Y.~Xie,
``Gravitational lensing by a black-bounce-Reissner\textendash{}Nordstr\"om spacetime,''
Eur. Phys. J. C \textbf{82}, no.5, 471 (2022)
%doi:10.1140/epjc/s10052-022-10441-7
%11 citations counted in INSPIRE as of 23 Sep 2023




%\cite{Bronnikov:2022bud}
\bibitem{Bronnikov:2022bud}
K.~A.~Bronnikov,
``Black bounces, wormholes, and partly phantom scalar fields,''
Phys. Rev. D \textbf{106}, no.6, 064029 (2022)
%doi:10.1103/PhysRevD.106.064029
[arXiv:2206.09227 [gr-qc]].
%19 citations counted in INSPIRE as of 23 Sep 2023

%\cite{Canate:2022gpy}
\bibitem{Canate:2022gpy}
P.~Ca\~nate,
``Black bounces as magnetically charged phantom regular black holes in Einstein-nonlinear electrodynamics gravity coupled to a self-interacting scalar field,''
Phys. Rev. D \textbf{106}, no.2, 024031 (2022)
%doi:10.1103/PhysRevD.106.024031
[arXiv:2202.02303 [gr-qc]].
%13 citations counted in INSPIRE as of 23 Sep 2023


%\cite{Lima:2023jtl}
\bibitem{Lima:2023jtl}
A.~Lima, G.~Alencar and D.~S\'aez-Chillon G\'omez,
``Regularizing Rotating Black Strings: a new black bounce solution,''
[arXiv:2307.07404 [gr-qc]].


%\cite{Lima:2023arg}
\bibitem{Lima:2023arg}
A.~Lima, G.~Alencar, R.~N.~Costa Filho and R.~R.~Landim,
``Charged black string bounce and its field source,''
Gen. Rel. Grav. \textbf{55}  no.10, 108 (2023)
%doi:10.1007/s10714-023-03156-x
[arXiv:2306.03029 [gr-qc]].


%\cite{Bronnikov:2023aya}
\bibitem{Bronnikov:2023aya}
K.~A.~Bronnikov, M.~E.~Rodrigues and M.~V.~de S.~Silva,
``Cylindrical black bounces and their field sources,''
Phys. Rev. D \textbf{108}  no.2, 024065 (2023)
%doi:10.1103/PhysRevD.108.024065
[arXiv:2305.19296 [gr-qc]].


%\cite{Pereira:2023lck}
\bibitem{Pereira:2023lck}
C.~F.~S.~Pereira, D.~C.~Rodrigues, J.~C.~Fabris and M.~E.~Rodrigues,
``Black-bounce solution in $k$-essence theories,''
[arXiv:2309.10963 [gr-qc]].
%0 citations counted in INSPIRE as of 23 Sep 2023


%\cite{Fitkevich:2022ior}
\bibitem{Fitkevich:2022ior}
M.~Fitkevich,
``Black bounces and remnants in dilaton gravity,''
Phys. Rev. D \textbf{105}, no.10, 106027 (2022)
%doi:10.1103/PhysRevD.105.106027
[arXiv:2202.00023 [gr-qc]].
%2 citations counted in INSPIRE as of 23 Sep 2023

%\cite{Junior:2022zxo}
\bibitem{Junior:2022zxo}
E.~L.~B.~Junior and M.~E.~Rodrigues,
``Black-bounce in $f(T)$ gravity,''
Gen. Rel. Grav. \textbf{55}, no.1, 8 (2023)
%doi:10.1007/s10714-022-03048-6
[arXiv:2203.03629 [gr-qc]].
%9 citations counted in INSPIRE as of 23 Sep 2023

%\cite{Junior:2023qaq}
\bibitem{Junior:2023qaq}
J.~T.~S.~S.~Junior and M.~E.~Rodrigues,
``Coincident $f(\mathbb {Q})$ gravity: black holes, regular black holes, and black bounces,''
Eur. Phys. J. C \textbf{83}, no.6, 475 (2023)
%doi:10.1140/epjc/s10052-023-11660-2
[arXiv:2306.04661 [gr-qc]].
%3 citations counted in INSPIRE as of 23 Sep 2023

%%%%%%%%%%%%%%%%%%%%%%%%%%%%%%%%%
%\cite{Ghosh:2021clx}
%\bibitem{Ghosh:2021clx}
%S.~G.~Ghosh and R.~K.~walia,
%``Rotating black holes in general relativity coupled to nonlinear electrodynamics,''
%Annals Phys. \textbf{434} (2021), 168619
%doi:10.1016/j.aop.2021.168619
%[arXiv:2109.13031 [gr-qc]].
%6 citations counted in INSPIRE as of 27 Sep 2023

%\cite{Bonanno:2000ep}
\bibitem{Bonanno:2000ep}
A.~Bonanno and M.~Reuter,
``Renormalization group improved black hole space-times,''
Phys. Rev. D \textbf{62} (2000), 043008
%doi:10.1103/PhysRevD.62.043008
[arXiv:hep-th/0002196 [hep-th]].
%470 citations counted in INSPIRE as of 01 Oct 2023

%\cite{Bronnikov:2021uta}
\bibitem{Bronnikov:2021uta}
K.~A.~Bronnikov and R.~K.~Walia,
``Field sources for Simpson-Visser spacetimes,''
Phys. Rev. D \textbf{105} (2022) no.4, 044039
%doi:10.1103/PhysRevD.105.044039
[arXiv:2112.13198 [gr-qc]].
%51 citations counted in INSPIRE as of 03 Feb 2024


%\cite{Do:2019txf}
\bibitem{Do:2019txf}
T.~Do, A.~Hees, A.~Ghez, G.~D.~Martinez, D.~S.~Chu, S.~Jia, S.~Sakai, J.~R.~Lu, A.~K.~Gautam and K.~K.~O'Neil, \textit{et al.}
``Relativistic redshift of the star S0-2 orbiting the Galactic center supermassive black hole,''
Science \textbf{365}  no.6454, 664-668 (2019)
%doi:10.1126/science.aav8137
[arXiv:1907.10731 [astro-ph.GA]].

%\cite{Nashed:2021pah}
\bibitem{Nashed:2021pah}
G.~G.~L.~Nashed and E.~N.~Saridakis,
``Stability of motion and thermodynamics in charged black holes in f(T) gravity,''
JCAP \textbf{05} (2022) no.05, 017
%doi:10.1088/1475-7516/2022/05/017
[arXiv:2111.06359 [gr-qc]].


%\cite{Kofinas:2015hla}
\bibitem{Kofinas:2015hla}
G.~Kofinas, E.~Papantonopoulos and E.~N.~Saridakis,
``Self-Gravitating Spherically Symmetric Solutions in Scalar-Torsion Theories,''
Phys. Rev. D \textbf{91} (2015) no.10, 104034
%doi:10.1103/PhysRevD.91.104034
[arXiv:1501.00365 [gr-qc]].

%%%%%%%%%%%%%%%%%%%%%%%%%%%%%%%%%

%%%%%%%%%%%%%%%%

\end{thebibliography}
\end{document}